\documentstyle[aps,prl,epsf,twocolumn,floats]{revtex}
\begin{document}
\draft
\twocolumn[\csname @twocolumnfalse\endcsname
\widetext

\title{
Critical local moment fluctuations in the Bose-Fermi Kondo model }
\author{Lijun Zhu and Qimiao Si}
\address{Department of Physics \& Astronomy, Rice University, Houston,
TX 77005--1892}

\maketitle

\begin{abstract}

We consider the critical properties of the Bose-Fermi Kondo model,
which describes a local moment simultaneously coupled 
to a conduction electron
band and a fluctuating magnetic field, i.e., a 
dissipative bath of vector bosons.
We carry out 
an $\epsilon-$expansion to higher than linear orders.
(Here $\epsilon$ is defined in terms of the power-law exponent
of the bosonic-bath spectral function.)
An unstable fixed point
is identified not only in the spin-isotropic case but also in the presence
of anisotropy. It marks the point where the weight of the Kondo resonance
has just gone to zero, and the local moment fluctuations are critical.
The exponent for the local spin susceptibility at this critical point
is found to be equal to $\epsilon$ in all cases. Our results imply that
a quantum phase transition of the ``locally critical'' type is 
a robust microscopic solution to Kondo lattices.
\end{abstract}
]

\narrowtext

\section{Introduction}
\label{sec:intro}

The interest in quantum criticality in strongly correlated metals
arises primarily because it provides a mechanism for non-Fermi
liquid behavior. While the problem appears to be important for 
a number of correlated electron systems -- including high
temperature superconductors -- the issues are particularly well-defined
in heavy fermion metals\cite{StewartRMP,vonLohneysen,Lonzarich,Steglich,%
Stewart,Flouquet,Thompson,deVisser,Maple,Andraka,Aronson,MacLaughlin}.
Here, many materials have been shown to explicitly display a magnetic
quantum critical point (QCP), including a growing list of 
stoichiometric or nearly stoichiometric ones
\cite{vonLohneysen,Lonzarich,Steglich,Stewart}.
A particularly striking puzzle has emerged from inelastic
neutron-scattering experiments\cite{Schroder2,Schroder1,Stockert,Stockert2}
and magnetization measurements\cite{Steglich,Schroder2,Schroder1}
in some of these (nearly) stoichiometric materials.
The dynamical spin susceptibility displays an $\omega/T$
scaling with a fractional energy/temperature exponent.
In addition, the same exponent is seen not only at the antiferromagnetic
ordering wavevector but also essentially everywhere else
in the Brillouin zone.

In a recent work\cite{QS_nature,QS_lcp,IJMPB,Review},
a ``locally-critical point'' is found in Kondo lattice systems.
This picture appears to explain the salient features of the
aforementioned experiments.
At a locally critical point, 
spatially extended critical fluctuations co-exist with spatially local
ones.
More microscopically, the criticality of the local Kondo physics
is ``embedded'' into that associated with the magnetic ordering
of the lattice: the divergence of the spatial correlation length
is accompanied by the destruction of the Kondo resonance.
The microscopic analysis\cite{QS_nature,QS_lcp}
was carried out within an extended 
dynamical mean field theory (EDMFT), in which the Kondo lattice system
is treated in terms of a single-impurity Bose-Fermi Kondo model
supplemented by a self-consistency condition. The impurity model
describes a local moment coupled at once to a conduction electron
band and a dissipative bath of vector bosons; the bosonic bath describes
a fluctuating magnetic field generated by the neighboring local moments.
The locally critical point is identified as a 
self-consistent solution
when the Bose-Fermi Kondo model is treated 
to the first order in $\epsilon$ within an $\epsilon-$expansion
renormalization group (RG) procedure.
Here $\epsilon \equiv 1-\gamma$, where $\gamma$ is the power-law exponent
of the spectrum of the dissipative bosonic bath (defined in
Eq. (\ref{dos-boson})).

There are several important questions that remain.
First, does a self-consistent locally-critical solution arise 
at higher orders in $\epsilon$? Second, what happens to the locally
critical point in spin-anisotropic situations? Such anisotropy
occurs in heavy fermions since the spin-orbit coupling is usually
strong in these systems.

These issues are addressed in the present paper. We carry out 
a detailed analysis of the Bose-Fermi Kondo model and show that
an unstable fixed point exists not only in the spin-isotropic case
but also in the presence of spin-anisotropy (both
xy and Ising cases).
In each case, this critical point
describes a continuous transition from a Kondo phase,
in which the local moment is quenched by the spins of the
conduction electrons, to a ``local-moment'' phase where there
is no Kondo resonance.
For the isotropic and xy cases, we calculate the local
spin susceptibility to the order $\epsilon ^ 2$ and,
in addition, we determine the associated critical
exponent, $\eta$ (defined in Eq. (\ref{eta-definition})),
to all orders in $\epsilon$. In the Ising case, 
a slightly different RG approach turns out to be useful.\cite{Smith1}
In all three cases, we find that the exponent $\eta$
at the unstable fixed point is
\begin{eqnarray}
\eta = \epsilon
\label{eta-result}
\end{eqnarray}
This result turns out to guarantee the existence of a self-consistent
solution of the locally-critical type
in the Kondo lattice systems.

We note in passing that the Bose-Fermi Kondo model is also of 
interest in other contexts. Historically, the model, in its
Ising version, was first studied in the context of an EDMFT
treatment of a spinless model in ref.~\onlinecite{Smith1},
using an $\epsilon-$expansion.
The spinful version was subsequently
considered, also through 
an $\epsilon-$expansion, in refs.~\onlinecite{Smith2,Sengupta}.
The Bose-only Kondo model was later extensively
analyzed
within a similar $\epsilon-$expansion in
ref.~\onlinecite{Vojta},
which went to higher orders in $\epsilon$ and established 
Eq. (\ref{eta-result}) for a stable fixed point of that model
to all orders.
The Bose-only Kondo model appears implicitly in a single-impurity
spin model first studied (using a large N approach)
in ref.~\onlinecite{Sachdev-Ye}; see also 
ref.~\onlinecite{GPS}. The Fermi-only Kondo model
is of course a standard textbook problem\cite{Hewson}.

The remainder of the paper is organized as follows.
In Section~\ref{sec:model}, we define the model and introduce
the formalism. Section~\ref{sec:rg} is devoted to the RG analysis,
to order $\epsilon^2$, in the isotropic case. Section~\ref{sec:suscep}
presents the explicit calculation of the local spin susceptibility
to order $\epsilon^2$, as well as the analysis of the associated 
critical exponent $\eta$ to all orders.
In Section~\ref{sec:aniso}, we consider
the effect of spin-anisotropy. Section~\ref{sec:lcp} discusses
the consequence of our results for the locally critical solution
in a Kondo lattice and Section~\ref{sec:sum} provides a brief
summary of our results. Some of the technical details are relegated to
Appendices~\ref{sec:appen-rg}, \ref{sec:appen-chiloc}
\ref{sec:appen-aniso},
and \ref{sec:appen-ising}.

\section{The model and formalism}
\label{sec:model}

The Bose-Fermi Kondo model is defined as follows 
\begin{eqnarray}
{\cal H}_{\text{BFK}}
=&& J ~{\bf S} \cdot {\bf s}_c
+ \sum_{p,\sigma} E_{p}~c_{p\sigma}^{\dagger}~ c_{p\sigma}
\nonumber\\ &&
+ \; g \sum_{p} {\bf S} \cdot
\left( \vec{\phi}_{p} + \vec{\phi}_{-p}^{\;\dagger} \right)
+ \sum_{p} w_{p}\,\vec{\phi}_{p}^{\;\dagger}\cdot \vec{\phi}_{p}.
\label{H-BFK}
\end{eqnarray}
A spin-${1 \over 2}$ local moment, ${\bf S}$, is coupled to 
both a fermionic bath ($c_{p\sigma}$), through the 
Kondo interaction $J$, and a dissipative vector-bosonic
bath ($\vec{\phi}_{p}$) with a coupling constant $g$.
$J$ is positive, i.e., antiferromagnetic, but $g$ can be either
positive or negative. (A sign change in $g$ can be absorbed
by a corresponding sign change in $\phi$.)
The spectral function of the bosonic bath is taken to have
a sublinear power-law dependence on energy, at sufficiently
low energies:
\begin{eqnarray}
\sum_{p}  [ \delta (\omega - w _p) - \delta (\omega + w _p)]
= (K_0^2 / \pi) |\omega|^{\gamma} {\rm sgn}\omega
\label{dos-boson}
\end{eqnarray}
for $|\omega| < \Lambda$. Here, the power-law exponent
\begin{eqnarray}
0<\gamma \equiv 1-\epsilon<1 .
\label{epsilon-defined}
\end{eqnarray}
The density of states of the conduction electron band
near the Fermi energy is taken to be a constant:
\begin{equation}
\sum_{p} \delta (\omega - E_{p}) = N_0.
\label{dos-fermion}
\end{equation}

The fluctuating magnetic field competes against the 
Kondo-singlet formation.
When the fluctuations are sufficiently slow such that $\epsilon > 0$,
$g$ is a relevant coupling in the RG sense and it can compete with
the marginally-relevant $J$ coupling. The physics of this competition
is amenable to an $\epsilon-$expansion\cite{Smith2,Sengupta,Smith1}.
This is in contrast to the Kondo fixed point, which occurs at
an infinite coupling and can only be 
addressed by strong-coupling methods\cite{Hewson}.

Following Smith and Si\cite{Smith2,QS_nature,QS_lcp},
we adopt the Abrikosov representation of the
spin in terms of pseudo-$f$-electrons\cite{Abrikosov},
\begin{equation}
{\bf S} = \sum_{\sigma \sigma'} f_{\sigma}^{\dagger}
{\vec{\tau}_{\sigma\sigma'} \over 2} f_{\sigma'} ,
\label{pseudo-fermion}
\end{equation}
where $\tau^{x,y,z}$ are the Pauli matrices.
In this representation, the Bose-Fermi Kondo Hamiltonian
takes the following form,
\begin{eqnarray}
\tilde{\cal H} && =~ {\cal H}_0 + 
{\cal H}_J
+{\cal H}_g , \nonumber\\
{\cal H}_0 && = ~ 
\sum_{\sigma } \lambda~f_{\sigma }^{\dagger}~f_{\sigma }
+ \sum_{p,\sigma} E_{p}~c_{p\sigma}^{\dagger}~ c_{p\sigma}
+ \sum_{p} w_{p}\,\vec{\phi}_{p}^{\;\dagger}\cdot \vec{\phi}_{p} ,
\nonumber\\[-1ex]
\label{H-BFK-tilde} \\[-1ex]
{\cal H}_{J} && = ~{J \over 4} \sum_{\sigma \sigma'} \sigma \sigma'
f_{\sigma}^{\dagger} f_{\sigma} c_{\sigma'}^{\dagger}
c_{\sigma'}
+{J \over 2} ( f_{\uparrow}^{\dagger} f_{\downarrow}
c_{\downarrow}^{\dagger}
c_{\uparrow} + \text{H.c.}),
\nonumber \\
{\cal H}_{g} && = ~{g \over 2} \sum_{\sigma } \sigma
f_{\sigma}^{\dagger} f_{\sigma} {\phi}^z
+{g \over \sqrt{2}} ( f_{\uparrow}^{\dagger} f_{\downarrow}
{\phi}^{-} + \text{H.c.}) ,
\nonumber
\end{eqnarray}
where $\lambda$ is an energy level for the $f-$electron that will be
set to $\infty$ at the end, $\sigma = \pm 1$, $\vec{\phi}
\equiv \sum_p (\vec{\phi}_p + \vec{\phi}_{-p}^{\;\dagger})$,
and ${\phi}^{\pm} = (\phi^x \pm \phi^y)/\sqrt{2}$.

To analyze the critical behavior of the Bose-Fermi Kondo model,
we carry out an RG procedure
based on a dimensional regularization
with a minimal subtraction (MS) of
poles\cite{Zinn-Justin,Brezin,Gross,tHooft}.
We define a renormalized field $f$
and a dimensionless coupling constant $g$ by
\begin{eqnarray}
 f_B &=& Z_f^{1/2}f,  \\
 g_B&=& g Z_f^{-1}Z_g \mu^{\epsilon/2},
\label{RF_def1}
\end{eqnarray}
where $Z_f$ is the wave-function renormalization factor for $f$ electrons,
$g_B$ is the bare coupling constant, $Z_g$ is a coupling constant 
renormalization of $g$, and $\mu$ is a renormalization energy scale.

The ``dimensional'' regularization  for the conduction electron density
of states is done by introducing an $\epsilon'$ through
\begin{equation}
\sum _p \delta(\omega-E_p)=N_0 |\omega|^{-\epsilon'},
\end{equation}
Accordingly, the dimensionless coupling constant $J$ is defined by
\begin{equation}
 J_{B}= J Z_f^{-1}Z_J \mu^{\epsilon'}.
\label{RF_def2}
\end{equation}
Note that $\epsilon '$ is introduced strictly for
the purpose of facilitating the MS analysis;
it is set to zero
at the final stage of the calculation.

\section{RG analysis}
\label{sec:rg}

In this section, we carry out the RG analysis to order $\epsilon^2$.

\subsection{RG flow equations}

It is known that, to the linear order of $\epsilon$, there
is a critical point at $(K_0 g^*)^2 = \epsilon/2$,
$(N_0J^*) = \epsilon/2$\cite{Smith2,Sengupta,QS_nature,QS_lcp}.
To obtain the corrections to the order $\epsilon^2$, we need
to calculate the self-energy and vertex corrections beyond the 
orders $J$ and $g^2$ and include also terms to the orders
$g^4$, $J^2$, and $g^2J$. The details of our calculation
are given in Appendix~\ref{sec:appen-rg}.

The renormalization factors are obtained as
\begin{eqnarray} 
Z_f =&& 1 - \frac {3}{4 \epsilon} (K_0g) ^2  
	- \frac 3{16\epsilon'} (N_0J)^2 \nonumber \\  
  &&- \frac {15}{32 \epsilon^2} (K_0g)^4 
  +\frac {3}{8 \epsilon} (K_0g)^4 \nonumber \\ 
\label{RF1} 
Z_g =&& 1 + \frac {1}{4 \epsilon} (K_0g)^2 + \frac {1}{16 \epsilon'}  
	(N_0J)^2 \nonumber \\ 
	&&+ \frac {9}{32 \epsilon^2} (K_0g)^4 
  -\frac {1}{8 \epsilon} (K_0g)^4,   \nonumber \\ 
 Z_J =&& 1 - \frac {1}{ \epsilon'} N_0J  \nonumber \\ 
  &&+ \frac {1}{4 \epsilon} (K_0g)^2 \nonumber 
  +\frac {1}{16 \epsilon'} (N_0J)^2  
  +\frac {1}{ \epsilon'^2} (N_0J)^2  \nonumber \\ 
  && - \left( \frac 1{\epsilon'(\epsilon+\epsilon')} +
 \frac {1}{4 \epsilon \epsilon'} \right) (K_0g)^2(N_0J) 
    \nonumber \\ 
 &&+ \frac {9}{32 \epsilon^2} (K_0g)^4 
  -\frac {1}{8 \epsilon} (K_0g)^4 . 
\label{RF2} 
\end{eqnarray} 
They can then be used
to determine the beta functions
for the coupling constants $g$ and $J$, defined as\cite{footnote}
\begin{eqnarray}
\beta(g) =&& \mu \frac{d g}{d \mu} \vert _{g_B, J_{B}}, \nonumber \\
 \beta(J) =&& \mu \frac{d J}{d \mu} \vert _{g_B, J_{B}}.
\label{beta-definition}
\end{eqnarray}
The results are given as follows
\begin{eqnarray}
\beta(g) =&& -g \left( \frac{\epsilon}{2}-(K_0g)^2 + (K_0g)^4  
	-\frac {(N_0J)^2}2\right),  \nonumber \\
\beta(J) =&&  -J \left( (N_0J) - \frac {(N_0J)^2}2
 \right) \nonumber \\
  &&   -J \left(-(K_0g)^2 + (K_0g)^4 \right). 
\label{rg-equations}
\end{eqnarray}

\subsection{The phase diagram}

\begin{figure}
\vspace{2ex}
\centering
\vbox{\epsfxsize=80mm\epsfbox{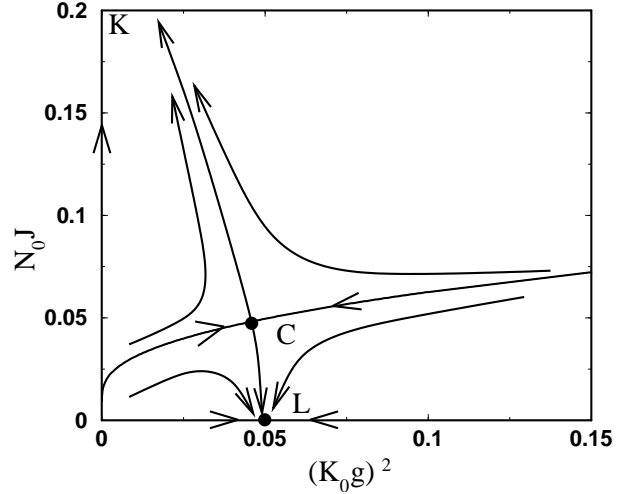}}
\vspace{3ex}
\caption{RG flows in $g-J$ plane when $\epsilon=0.1$.
K (Kondo) and L (local-moment) denote the two stable
fixed points. The separatrix is determined numerically.
C denotes the unstable fixed point (the critical point).}
\label{fig:RGflow_e01}
\end{figure}

The RG equations~(\ref{rg-equations}) yield several fixed points.
There are two stable ones.
The fixed point located at $g^*=0$ and large $J^*$ 
(``K'' in Fig.~\ref{fig:RGflow_e01}), as usual, specifies
the Kondo phase\cite{footnote2}. Here, the local moment
is quenched by the spins of the conduction electrons,
leading to the development of a Kondo resonance.

Another stable fixed point (``L'' in Fig.~\ref{fig:RGflow_e01})
is located at 
\begin{eqnarray}
{(K_0g^*}) ^2 &=& \frac{\epsilon}{2}+\frac{1}{4}\epsilon^2 +
O(\epsilon^3) \nonumber \\
(N_0J^*)  &=& 0
\label{fixed-point-stable}
\end{eqnarray}
We will call it a ``local-moment'' fixed point, emphasizing
the fact that it describes a phase in which no Kondo resonance
arises. The dynamics in the local-moment phase is controlled
by the coupling of the local moment to the dissipative bosonic
bath alone.

A separatrix specifies the boundary of the domains of attractions
for these two phases in the J-g parameter space. Lying on this
separatrix is an unstable fixed point, or a critical point
(``C'' in Fig.~\ref{fig:RGflow_e01}). It is located  at 
\begin{eqnarray}
{(K_0g^*}) ^2 &=& \frac{\epsilon}{2}+\frac{1}{8}\epsilon^2 
+O(\epsilon^3) \nonumber \\
(N_0J^*)  &=& \frac{\epsilon}{2} + O(\epsilon^3)
\label{fixed-point}
\end{eqnarray}
The critical point marks the point where the spectral weight
of the Kondo resonance goes to zero. It captures the 
competition between the Kondo coupling of the local moment to
the conduction electrons on the one hand, and its coupling
to the fluctuating magnetic field on the other.

The RG flow can be solved numerically. 
In Fig.~\ref{fig:RGflow_e01}, we give the result
for $\epsilon=0.1$.

\section{The local dynamical spin susceptibility}
\label{sec:suscep}

We now turn to the calculation of the local susceptibility.

\subsection{Critical susceptibility to order $\epsilon^2$}

Consider the local spin susceptibility to order $\epsilon^2$.
We first calculate the bare spin susceptibility,
i.e., the autocorrelation function of the unrenormalized local
spin operator specified by Eq.~(\ref{pseudo-fermion}),
\begin{eqnarray}
\chi_B (\tau) && \equiv { 1 \over 2} 
\langle T_{\tau} S^- (\tau) S^+ (0) \rangle
 = \lim_{\lambda \rightarrow \infty} {1 \over 2}~ e^{\beta \lambda}~
 \tilde {\chi}_B(\tau) , \nonumber \\
\tilde {\chi}_B
(\tau) &&\equiv {1 \over 2} \langle T_{\tau}
f_{\downarrow}^{\dagger}(\tau) f_{\uparrow} (\tau)
f_{\uparrow}^{\dagger}(0) f_{\downarrow} (0) \rangle ,
\label{chi_loc_define}
\end{eqnarray}
where $S^{\pm}\equiv (S^x \pm iS^y)$.
(For notational simplicity, we've dropped the 
subscript $B$ in the operators used in Eq.~(\ref{chi_loc_define}).)

Correspondingly, we define a spin renormalization factor $Z$ which relates
the renormalized spin susceptibility, $\chi$, to the bare susceptibility,
\begin{equation}
\chi(\tau) = \frac 1Z \chi_B(\tau) ,
\label{Z-defined}
\end{equation}
The calculational details are given
in Appendix \ref{sec:appen-chiloc}. From 
Eq.~(\ref{RFZ_inverse}), $Z$ is given by
\begin{equation}
Z = 1 -  \frac{2(K_0g)^2}{\epsilon}
+\frac{(K_0g)^4}{\epsilon}
  -\frac {(N_0J)^2}{2\epsilon'}.
\label{RF3}
\end{equation}

At the critical point, the local spin susceptibility is expected to have
a power-law form:
\begin{eqnarray}
\chi(\tau) \approx A_{\epsilon} \left( \frac {\tau_0}{| \tau | }
\right )^{\eta},
\label{eta-definition}
\end{eqnarray}
for $|\tau| \gg \tau_0$, where $\tau_0 = 1 /\Lambda $ is a cut-off scale.
Here, $\eta$ is the anomalous dimension. 
It can be calculated from the renormalization factor $Z$, 
\begin{eqnarray}
 \eta =&& \mu \frac{d \ln Z}{d \mu} \vert _{g=g^*, J=J^*} \nonumber \\
       =&&\left( \beta(g)\frac { \partial \ln Z}{\partial g}
  + \beta(J)\frac { \partial \ln Z}{\partial J} \right) \vert_{g=g^*, J=J^*}
   \nonumber \\ 
  =&& 2{(K_0g^*)}^2 - 2{(K_0g^*)}^4 + {(N_0J^*)}^2 \nonumber \\
   =&& \epsilon.
\label{eta-expression}
\end{eqnarray}
It is striking that, 
although the order $\epsilon^2$ corrections are present 
both in the beta functions and in the critical coupling constants,
they cancel out in the critical exponent $\eta$, leaving only a 
non-zero linear term.

The critical amplitude is given in Appendix~\ref{sec:appen-chiloc}.

For completeness, we also briefly discuss the local spin
susceptibility at the local-moment fixed point.
We assume that the susceptibility here also has 
a power-law form:
\begin{equation}
 \chi(\tau) \sim A_{\epsilon}^L  \left( 
\frac {\tau_0}{| \tau |} \right )^{\eta_L} ,
\end{equation}
with an anomalous dimension $\eta_L$. $\eta_L$ has the same dependence
on $g^*, J^*$ as in Eq.~(\ref{eta-expression}). Even though
the coupling
constants $g^*, J^*$ now take the values specified 
by Eq.~(\ref{fixed-point-stable}),
it turns out that the $\epsilon^2$ terms again cancel with each other
leaving 
\begin{equation}
 \eta_L = \epsilon .
\end{equation}
The critical amplitude is also given in
Appendix~\ref{sec:appen-chiloc}.

\subsection{Critical exponent to all orders in $\epsilon$}
\label{sec:suscep2}

In this section, we show that $\eta=\epsilon$ given in 
Eq. (\ref{eta-expression}) is in fact exact to all orders.

First, we explore the reason of the cancellation of the $\epsilon^2$
terms in $\eta$ seen in Eq. (\ref{eta-expression}).
Using the equations (\ref{RF2}), (\ref{RF3}), 
we can easily verify that
\begin{equation}
Z^{-1}=(Z_f^{-1}Z_g)^2
\label{RF_relation}
\end{equation}
both to linear and second orders in $\epsilon$.
Combining Eq. (\ref{RF_relation}) with Eq. (\ref{RF_def1}) lead to
\begin{equation}
g_B= g Z^{-1/2} \mu^{\epsilon/2}.
\label{gb-g-Z}
\end{equation}
Differentiating the logarithm of both sides with respect to $\ln \mu$,
keeping $g_B$ and $J_B$ fixed, we have
\begin{equation}
\eta = \epsilon + {2 \beta(g) \over g} \vert_{g=g^*, J=J^*}.
\label{eta-epsilon}
\end{equation}
At the fixed point, $\beta(g)=0$, which gives
\begin{equation}
\eta = \epsilon. 
\label{critical-exp}
\end{equation}

There is also a consistency check for our result.
We can start from Eq. (\ref{RF_def2}). Combining this equation
with Eq. (\ref{RF_relation}) lead to
\begin{eqnarray}
J_{B}= J Z^{-1/2}(Z_J/Z_g).
\label{J-Z-Zj-Zg}
\end{eqnarray}
(We have set $\epsilon'=0$.) 
Again, differentiating the logarithm of both sides with respect
to $\ln \mu$, and again keeping $g_B$ and $J_B$ fixed, we end up with
\begin{equation}
\eta = 2 {{ d \ln (Z_J/Z_g) } \over {d \ln \mu}}
\vert_{g=g^*, J=J^*}.
\end{equation}
To be consistent, the RHS should be equal to $\epsilon$.
This is indeed verified to both linear and quadratic orders
in $\epsilon$.

We now proceed to higher order contributions.
We show in Appendix~\ref{sec:appen-chiloc} that
Eq.~(\ref{RF_relation}) is valid to all orders.
Eq. (\ref{eta-epsilon}) is then valid to all orders in $\epsilon$.
By extension,
$\eta$ equals to  $\epsilon$ to all orders in $\epsilon$.
Our reasoning basically parallels that of ref.~\onlinecite{Vojta}
for the stable fixed point in the Bose-only Kondo problem.

\section{Anisotropic cases (xy and Ising)}
\label{sec:aniso}

So far, we have discussed the spin-isotropic case.
We now turn to the effect of anisotropy in spin-space.

\subsection{RG equations}
\label{sec:aniso-rg}

We introduce separate parameters for the transverse and longitudinal
spin couplings as follows:
\begin{eqnarray}
{\cal H}_{J} && = {J_{z} \over 4} \sum_{\sigma \sigma'} \sigma \sigma'
f_{\sigma}^{\dagger} f_{\sigma} c_{\sigma'}^{\dagger}
c_{\sigma'}
+{J_{\perp} \over 2} ( f_{\uparrow}^{\dagger} f_{\downarrow}
c_{\downarrow}^{\dagger}
c_{\uparrow} + \text{H.c.}),
\nonumber \\[-1ex]
\label{coupling_aniso} \\[-1ex]
{\cal H}_{g} && = {g_z \over 2} \sum_{\sigma } \sigma
f_{\sigma}^{\dagger} f_{\sigma} {\phi}^z
+{g_{\perp} \over \sqrt{2}} ( f_{\uparrow}^{\dagger} f_{\downarrow}
{\phi}^{-} + \text{H.c.}) ,
\nonumber
\end{eqnarray}

\begin{figure}
\vspace{2ex}
\centering
\vbox{\epsfxsize=65mm\epsfbox{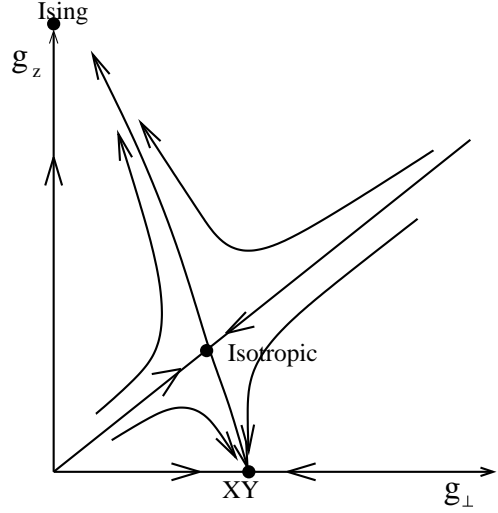}}
\vspace{3ex}
\caption{Schematic RG flow diagram for anisotropic $g$ coupling when
$J=0$. The Ising fixed-point and the xy fixed-point are stable while
the isotropic fixed point is unstable.}
\label{fig:Aniso_g}
\end{figure}

The RG procedure parallels that for the isotropic case,
except that we need to differentiate between the longitudinal and transverse
couplings in all the contributions, and associate different renormalization
factors for the different types of vertices:
\begin{eqnarray}
g_{i}^B &=& g_i Z_f^{-1}Z_{g_i} \mu^{\epsilon/2} \nonumber \\
J_{i}^B &=& J_{i} Z_f^{-1}Z_{J_i} \mu^{\epsilon'},
\label{DefRF}
\end{eqnarray}
where $i=\perp, z$. Up to the order of our interest, the
results for the renormalization factors are given in
Appendix~\ref{sec:appen-aniso}.
The resulting RG equations are given as follows,
\begin{eqnarray}
\beta (g_{\perp}) &=& - \frac{\epsilon}{2} g_{\perp}  +
    \frac{(K_0 g_{\perp})^2+(K_0 g_z)^2}2 g_{\perp}  \nonumber \\
  &-&  \frac{(K_0 g_{\perp})^2 (K_0 g_z)^2+(K_0 g_{\perp})^4}{2}
    g_{\perp} \nonumber \\
  &+&\frac{(N_0J_{\perp})^2+(N_0J_z)^2}{4} g_{\perp} 
\label{beta_gperp_aniso}
\end{eqnarray}
\begin{eqnarray}
\beta (g_{z}) &=& - \frac{\epsilon}{2}
g_{z}  +
    (K_0 g_{\perp})^2 g_{z}  \nonumber \\
  &-&  (K_0 g_{\perp})^2 (K_0 g_z)^2
    g_{z} \nonumber \\
  &+&\frac{(N_0J_{\perp})^2}{2} g_{z} 
\label{beta_gz_aniso}
\end{eqnarray}
\begin{eqnarray}
\beta (J_{\perp}) &=&
   \frac{(K_0 g_{\perp})^2+(K_0 g_z)^2}2 J_{\perp}  \nonumber \\
  &-&  \frac{(K_0 g_{\perp})^2 (K_0 g_z)^2+(K_0 g_{\perp})^4}{2}
    J_{\perp} \nonumber \\
  &-&   (N_0J_z) J_{\perp} +
    \frac{(N_0J_{\perp})^2+(N_0J_z)^2}{4} J_{\perp} 
\label{beta_Jperp_aniso}
\end{eqnarray}
\begin{eqnarray}
\beta (J_{z}) &=&
    (K_0 g_{\perp})^2 J_{z}  \nonumber \\
  &-&  (K_0 g_{\perp})^2 (K_0 g_z)^2
    J_{z} \nonumber \\
  &-& (N_0J_{\perp}) J_{\perp} +
    \frac{(N_0J_{\perp})^2}{2} J_{z}
\label{beta_Jz_aniso}
\end{eqnarray}
Consider first the $J=0$ case, i.e. the Bose-only Kondo model.
The RG flow is given in Fig.~\ref{fig:Aniso_g}.
There are two more fixed points in addition to the isotropic one.
The xy fixed point is located at
\begin{eqnarray}
(K_0g_{\perp}^*)^2 && = \epsilon + \epsilon^2+ O(\epsilon^3)
\nonumber\\
(K_0g_{z}^*) && = 0.
\label{fp-xy-stable}
\end{eqnarray}
While the Ising fixed point is nominally located at $g_{\perp}^*=0$ and 
$g_z^* = \infty$. 
Both the isotropic fixed point and the xy fixed point
are accessible by the $\epsilon-$expansion. The Ising fixed point,
on the other hand, is beyond 
the reach of the perturbative RG scheme.

The $J-$coupling can lead away from the ``local-moment''
fixed point not only
in the isotropic case, but also
in the xy and Ising cases. The isotropic case has already been discussed
in the previous sections. We now turn to the xy and Ising cases, respectively.

\subsection{Critical behavior in the xy case}
\label{sec:xy}

\begin{figure}
\vspace{2ex}
\centering
\vbox{\epsfxsize=80mm\epsfbox{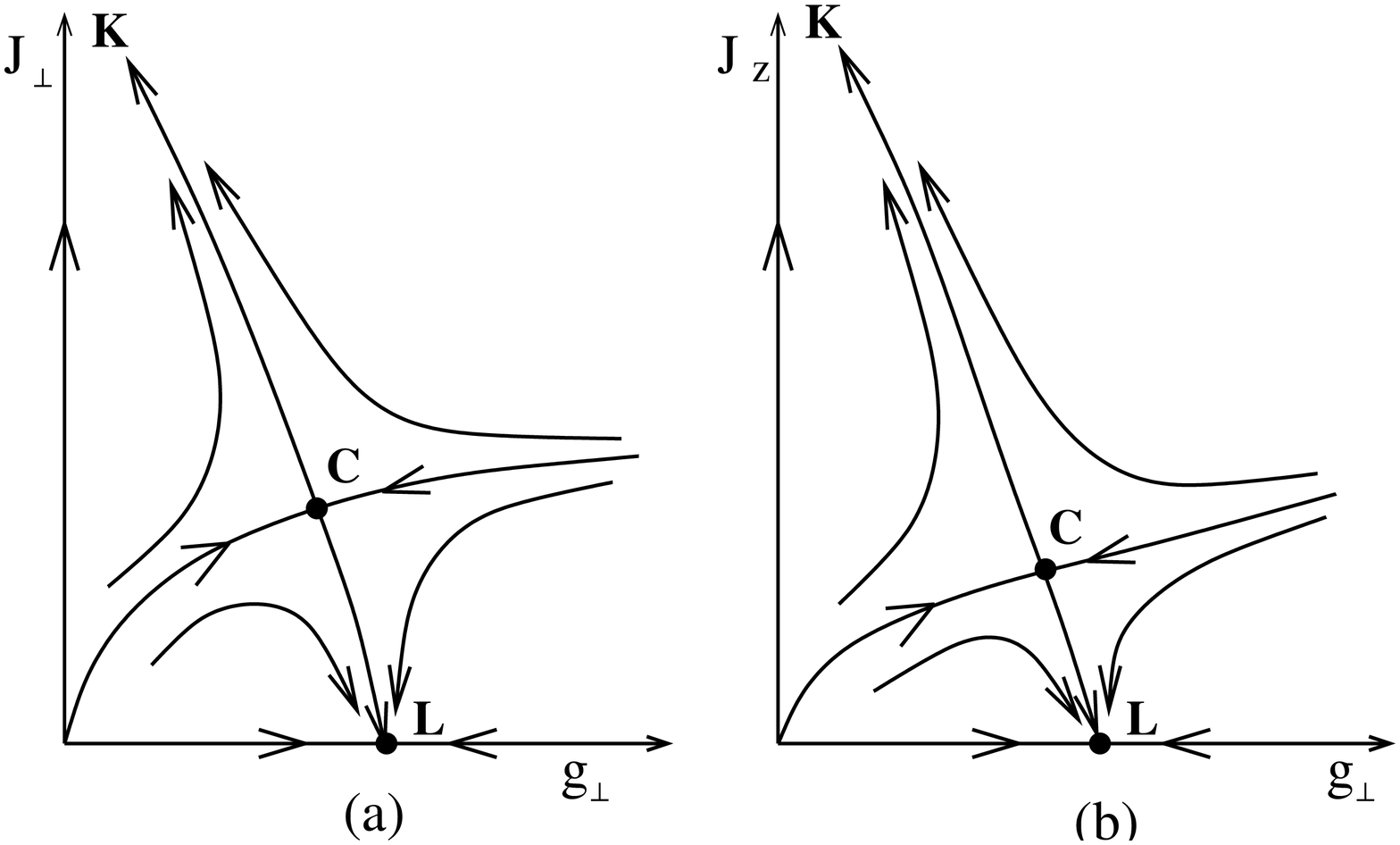}}
\vspace{3ex}
\caption{Schematic RG flow diagram for the xy case. Here we show 
the projection of the RG flow in the three dimensional parameter 
space ($J_{\perp}$, $J_z$, $g_{\perp}$) onto 
a) the $J_{\perp}-g_{\perp}$ plane and b) the $J_{z}-g_{\perp}$ plane.}
\label{fig:XY_RGflow}
\end{figure}

We will now set $g_{z}=0$ in Eqs. (\ref{beta_gperp_aniso})-
(\ref{beta_Jz_aniso}).
[Eq. (\ref{beta_gz_aniso}) implies that $g_z$ will stay at zero
under the RG transformation when its initial value vanishes.]
An unstable fixed point occurs at
\begin{eqnarray}
(K_0g_{\perp}^*)^2 && = \epsilon + \frac{5\epsilon^2}8 + O(\epsilon^3)
\nonumber\\
N_0J_{\perp}^* && = {\epsilon \over \sqrt{2}} +
 \frac {7}{16\sqrt{2}}\epsilon^2+ O(\epsilon^3) 
 \nonumber \\
N_0J_{z}^* && = {\epsilon\over 2} + O(\epsilon^3) 
\label{fp-xy}
\end{eqnarray}
Note that $g_z$ remains irrelevant near this fixed point,
establishing the consistency of our analysis.
The RG flows are shown in Fig.~\ref{fig:XY_RGflow}.

The perturbative correction to the local spin susceptibility can be
calculated as in the isotropic case. The most general expression for
the susceptibility as well as its renormalization factor $Z$
are given in Appendix~\ref{sec:appen-aniso}.
The corresponding expression for $\eta$ is also given there,
in Eq.~(\ref{crit_exp_ansio_xy}).
Using the values of the coupling constants at this fixed point,
Eq.~(\ref{fp-xy}), 
we find that $\eta$ remains to be $\epsilon$
to order $\epsilon^2$.

The same argument made in
Appendix~\ref{sec:appen-chiloc}
for the isotropic case carries
through here for the xy-case as well, resulting in
\begin{equation}
Z^{-1}=( Z_f ^{-1} Z_{g_{\perp}})^2,
\label{RF_relation_aniso}
\end{equation}
to infinite orders in $\epsilon$.
So $\eta=\epsilon$ is again valid to all orders in $\epsilon$.

We now again briefly examine the local spin susceptibility at
the local-moment fixed point in this case, where $J_{\perp}^*=J_z^*=0$,
and
\begin{equation}
(K_0g_{\perp}^*)^2=\epsilon+\epsilon^2+O(\epsilon^3).
\label{fp-xy-lm}
\end{equation}
The anomalous dimension $\eta_L$ has the same expression as in 
Eq.~(\ref{crit_exp_ansio_xy_lm}). Substituting the values
of the stable fixed point, we find that $\eta_L$ is equal
to $\epsilon$, just as in the isotropic case.
 
\subsection{Critical behavior in the Ising case}
\label{sec:ising}

None of the non-trivial fixed points 
in the Ising case is within the reach of the
perturbative RG scheme. For the unstable fixed point, for instance,
setting $g_{\perp}=0$ in Eqs. 
(\ref{beta_gperp_aniso}-\ref{beta_Jz_aniso})
will yield
a $(N_0J_{\perp}^*)^2 \sim \epsilon$ that is still small,
but $J_z^*$ and $g_z^*$ that are of order unity. The latter violates
our starting assumption.

\begin{figure}
\vspace{2ex}
\centering
\vbox{\epsfxsize=80mm\epsfbox{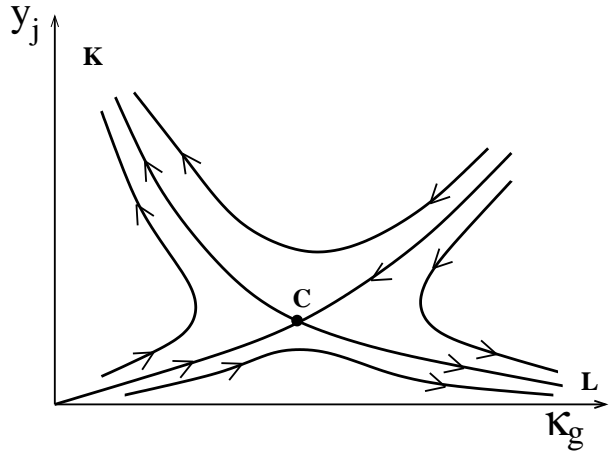}}
\vspace{3ex}
\caption{Schematic RG flow diagram for the Ising case, projected onto
the plane spanned by $y_j-\kappa_g$ plane.}
\label{fig:Ising_RGflow}
\end{figure}

This problem, however, has already been studied in
ref.~\onlinecite{Smith1}. A finite $J_z$ and $g_z$ can be handled by
introducing a so-called ``kink-gas'' representation. 
In this representation, the unstable fixed point is still
accessible by an $\epsilon-$expansion. The calculation is outlined 
in Appendix~\ref{sec:appen-ising}. The RG equations are no longer
constructed in terms of the bare couplings. Instead, they are
given
in terms of the stiffness constants $\kappa_j$, $\kappa_g$, and 
the fugacity $y_j$, whose initial values
are  specified by the bare parameters as follows,
\begin{eqnarray}
\kappa_j^B &&  = [1 - {1 \over \pi} \tan^{-1}({\pi \over 4}N_0J_z^B) ]^2
\nonumber\\
\kappa_g^B && =
\frac{\Gamma(\gamma)}{4} \tau_0^{1 - \gamma} (K_0g_z^B)^2
\nonumber\\
y_j^B && = {{N_0J_{\perp}^B} \over 2}
\label{RG-charges-ising}
\end{eqnarray}
The RG flow projected onto the $y_j-\kappa_g$ plane is shown in
Fig.~\ref{fig:Ising_RGflow}. In the vicinity of the unstable fixed point,
$\kappa_j$ is irrelevant.

The exponent $\eta$ at the unstable fixed point is calculated in
Appendix~\ref{sec:appen-ising} to the linear order in $\epsilon$.
Again, it is found that
\begin{eqnarray}
\eta = \epsilon
\label{eta-Ising}
\end{eqnarray}
In the same Appendix, it is also shown that $\eta$ is in fact 
the same as that for the critical point of a classical
ferromagnetic Ising chain with a long-range interaction
that decays in distance in terms of a power-law exponent $2-\epsilon$.
For this problem, it has 
long been
held that Eq. (\ref{eta-Ising})
is in fact exact\cite{Fisher,Suzuki}.

In passing, we note that this analogy also allows us to state
what happens to the local susceptibility at the local-moment
fixed point shown in Fig.~(\ref{fig:Ising_RGflow}).
Here, $\chi(\tau)$ picks up a constant ($\tau-$independent) piece.
At the same time,
the ``connected'' susceptibility (defined such that the constant piece
does not appear) decays with an exponent $\eta_L=2-\epsilon$\cite{Spohn}.

\section{Locally critical point of the Kondo lattice}
\label{sec:lcp}

We now discuss the consequences of our results for the locally critical
point of the Kondo lattices. A self-consistent microscopic treatment
of the Kondo lattice model has been presented in detail
elsewhere\cite{QS_nature,QS_lcp}. This was carried out 
within the EDMFT approach developed in\cite{Smith1,Smith3,Chitra}.
(This approach is a generalization of the dynamical mean
field theory\cite{Georges} such that magnetic fluctuations are also
taken into account.) The self-consistent equation reads as follows,
\begin{eqnarray}
< \chi ( {\bf q},\omega ) >_{\bf q} = \chi_{\rm loc} (\omega),
\label{self-consistent}
\end{eqnarray}
Here, the lattice dynamical spin susceptibility has the form
\begin{eqnarray}
\chi ({\bf q},\omega) = { 1 \over {M(\omega ) + I_{\bf q}}},
\label{chi-q-omega}
\end{eqnarray}
where $ M(\omega) $ is the ``spin self-energy'' and $I_{\bf q}$ describes
the exchange (RKKY) interaction between the local moments.
Eq. (\ref{self-consistent}) is simply a statement of translational
invariance, as it equates the on-site spin susceptibility of the
lattice system (LHS) to the susceptibility of any local moment (RHS).
The locally critical solution arises when
$\chi_{\rm loc}(\omega)$ is singular
-- signaling the destruction of the Kondo resonance as discussed
in previous sections -- at the point where the peak susceptibility
$\chi ({\bf Q}_{AF}, \omega)$ 
just diverges.
It is shown in refs.~\onlinecite{QS_nature,QS_lcp} that
one requirement for a locally critical point is that the
magnetic fluctuations are two-dimensional. 
This can be easily seen from 
Eqs.~(\ref{self-consistent},\ref{chi-q-omega}).
The fact that the peak susceptibility $\chi ({\bf Q}_{AF}, \omega)$ 
diverges 
means that $M(0)=-I_{{\bf Q}_{AF}}$.
Recognizing that for ${\bf q} \sim {{\bf Q}_{AF}}$,
the interaction has the generic form,
$I_{\bf q} = I_{ {\bf Q}_{AF}} 
+ a ({\bf q}- {\bf Q}_{AF} )^2$, we have
\begin{eqnarray}
\chi ({\bf q} \sim {\bf Q}_{AF},\omega \rightarrow 0) = { 1 \over {
a ({\bf q}- {\bf Q}_{AF} )^2}}
\label{chi-q-near-Q}
\end{eqnarray}
Now, the fact that the local susceptibility, 
$\chi_{\rm loc}(\omega \rightarrow 0)$,
is singular demands that the ${\bf q}-$averaging of the LHS of Eq. 
(\ref{self-consistent}) should
diverge.
In other words, integrating the RHS of Eq. (\ref{chi-q-near-Q})
over ${\bf q}$ should yield a singular result.
Two-dimensionality provides the phase space for this.

This condition, however, is not sufficient. It turns out that a second 
condition must be met for the existence of such a locally-critical
solution. The condition is precisely Eq. (\ref{eta-result}).
The origin of this second condition is somewhat subtle,
but has to do with the fact that $M(\omega)$ and $\chi_{\rm loc}$,
in addition to being related through
Eqs. (\ref{self-consistent},\ref{chi-q-omega}), must also satisfy
a Dyson-like equation of the (self-consistent) Bose-Fermi Kondo
model. The details can be found in ref.~\onlinecite{QS_lcp}.
By proving that Eq. (\ref{eta-result}) is valid to all orders in
$\epsilon$, and also in spin-anisotropic cases, we have then
established that the locally critical point is indeed a robust result
within these microscopic considerations. (The arguments for the 
robustness of the locally critical point beyond the microscopic approaches
can be found in refs.~\onlinecite{QS_nature,QS_lcp}.)

We stress that 
a key element 
of the locally critical point of the 
Kondo lattice is that, the criticality of the local degrees of freedom
is embedded in the criticality associated with the long-range ordering
in the lattice. In other words, the point where a local energy scale
turns to zero coincides with where the spatial correlation length just
diverges. 
This is very different from the self-consistent spin liquid solutions 
discussed in other contexts\cite{Sachdev-Ye,GPS},
which correspond to a phase instead of a critical point. This difference
is also reflected in the fact that stabilizing the locally critical
solution in Kondo lattices requires two-dimensional magnetic fluctuations.

\section{Summary}
\label{sec:sum}

In short, we have carried out a detailed analysis of the Bose-Fermi Kondo
model both when the spin-rotational invariance is satisfied 
(isotropic) and when it is broken (xy and Ising). 
In each case, we have identified a critical point that separates
a Kondo phase, where the local moment is quenched by the spins of
the conduction electrons, and a local moment phase where there exists
no Kondo resonance. This unstable fixed point marks the point where
the spectral weight associated with the Kondo resonance
has just gone down to zero.

In all three cases, we find that the exponent for the local
spin susceptibility at the unstable fixed point is equal to 
$\epsilon$. 
We note that the three cases correspond to three different one-dimensional
statistical mechanical problems.
It is remarkable that, the susceptibility exponent at the unstable
fixed point is insensitive to these differences. 
However, the 
same cannot
be said about the stable fixed points: here, the susceptibility
exponent for the Ising case is very different from the
isotropic and xy cases.

Our results have important consequences for the locally critical behavior
in Kondo lattices. In particular, it guarantees that a quantum phase
transition of a locally-critical type\cite{QS_nature,QS_lcp,IJMPB}
is a robust microscopic solution to the Kondo lattices: it arises
to all orders in $\epsilon$ and, in addition, not only when the system is
spin-rotationally invariant but also in spin-anisotropic situations.

\acknowledgments

This work has been supported in part by TCSUH, the Robert A. Welch Foundation,
and NSF Grant No.\ DMR-0090071. We would also like to acknowledge
the support of Argonne National Laboratory, the University of Chicago,
and the University of Illinois at Urbana-Champaign 
during our leave of absence from Rice University.
This work was briefly reported in ref.~\onlinecite{aps}. We have
recently learned\cite{Zarand}
that G. Zarand and E. Demler have also analyzed this
problem, with results which are similar to ours.

\vskip 0.2 in
\appendix
\section{RG procedure to order $\epsilon^2$} 
\label{sec:appen-rg} 
 
In this appendix, we give the details of the RG analysis.
We adopt the dimensional regularization and 
minimal subtraction scheme 
\cite{Zinn-Justin,Brezin,Gross,tHooft}.

We first carry out a renormalized perturbation calculation.
The quantities of interest are the $J-$ and $g-$ vertices:
\begin{eqnarray} 
\Gamma_J(\omega+\lambda) && \equiv J \mu^{\epsilon'}\gamma_J(\omega+\lambda) 
  \nonumber\\
\Gamma_g(\omega+\lambda) && \equiv g \mu^{\epsilon/2}\gamma_g (\omega+\lambda)
\end{eqnarray}
as well as the $f-$electron self-energy, $\Sigma_f(\omega+\lambda)$.
\begin{equation}
G^{-1}(\omega+\lambda)=\omega  -\Sigma_f(\omega+\lambda)
=Z_f G_B^{-1}(\omega+\lambda)
\end{equation}
Within the renormalized perturbation calculation,
the bare parameters in Eq.~(\ref{H-BFK-tilde})
are replaced with the corresponding renormalized ones.
We end up with a Hamiltonian containing 
one part that has the form of Eq.~(\ref{H-BFK})
(the tree level Hamiltonian)
and another part which describes the counterterms:
\begin{eqnarray} 
{\cal H}_{ct}=&&\sum_{\sigma }
(Z_{f}-1)\lambda~f_{\sigma }^{\dagger}~f_{\sigma }
+(Z_{J}-1)J\mu ^{\epsilon ^{\prime }}{\bf S}\cdot
{\bf s}_{c} \nonumber \\
&&+(Z_{g}-1)g\mu ^{\epsilon /2}{\bf S}\cdot \vec{\phi},
\end{eqnarray}
There are then two types of perturbative diagrams,
one coming from the tree level Hamiltonian,
which we will call direct perturbative contributions below,
and the other from the counterterms.

The singular contributions are kept track of through poles
as a function of $\epsilon$ and $\epsilon'$.
By demanding that such poles are ``minimally'' removed
so that the renormalized quantities $\Gamma_g$, $\Gamma_J$, 
and $G_f$ are regular (as a function of $\epsilon$ and $\epsilon'$),
we fix the wavefunction and vertex 
renormalization factors $Z_f$,
$Z_J$, and $Z_g$ order by order in the coupling constants.
From these renormalization factors, we can determine the 
RG beta functions as well as the anomalous dimensions.
Note that what we are carrying through is a double expansion
in $\epsilon$ and $\epsilon'$.

In the following, we separate our discussions according to the order
in $J$ and $g$.
 
\subsection{First order result} 
To the first order of perturbation, there are only 
corrections to the $J-$coupling from processes that 
involve the $J-$coupling alone.
The diagrams are specified by Figs. 5(a)(b) in  
Ref.~\onlinecite{QS_lcp}.  The result is given by 
\begin{eqnarray} 
\gamma_J^{(1)}(\omega+\lambda)=&&\frac{N_0J}{\epsilon'}
({\mu \over - \omega})^{\epsilon'} \nonumber \\
=&& \frac{N_0J}{\epsilon'}+ {N_0J} \ln ({\mu \over - \omega}) + \ldots
\end{eqnarray} 
Adding the counterterm $(Z_J-1)$ and demanding that the pole terms cancel,
we find that
\begin{equation} 
(Z_J-1)^{(1)} = - \frac {1}{ \epsilon'} (N_0J). 
\end{equation} 

\subsection{Second order result} 
\begin{figure} 
\vspace{2ex} 
\centering 
\vbox{\epsfxsize=65mm\epsfbox{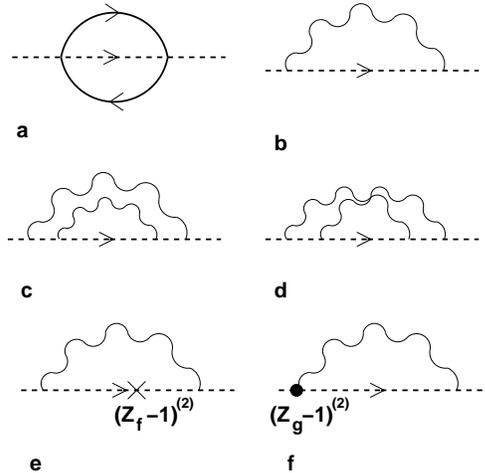}} 
\vspace{3ex} 
\caption{Self-energy diagrams for the $f-$electrons. The dashed, solid
and wavy lines denote the propagators of the $f$-electrons, conduction
electrons and  vector bosons, respectively. The cross in (e) and dot
in (f) denote the counterterm $Z_f-1$ and $Z_g-1$, respectively. } 
\label{fig:f-selfenergy} 
\end{figure} 
 
To the order $g^2$ and $J^2$, the contributions
to  the $f-$electron self-energy are shown in 
Figs.~\ref{fig:f-selfenergy}a and \ref{fig:f-selfenergy}b and they
yield the following expression:
\begin{eqnarray} 
\Sigma_f^{(2)}(\omega+\lambda) =&& 
 - \omega \frac 3{4\epsilon} (K_0g)^2   
({\mu \over - \omega})^{\epsilon} 
  \nonumber \\ 
 && - \omega \frac 3{16\epsilon '}(N_0J)^2 
 ({\mu \over - \omega})^{2\epsilon'} , 
 \label{selfenergy2} 
\end{eqnarray} 
By demanding that these poles of the self-energy
be cancelled by the 
counterterm contribution $-(Z_f-1)\omega$, we fix
\begin{equation} 
(Z_f-1)^{(2)} = - \frac {3}{4 \epsilon} (K_0g) ^2 
 - \frac {3}{16 \epsilon'} (N_0J)^2. 
\end{equation} 
 
To the same order, the corrections to the coupling constant $g$
are specified by Fig. 6 of Ref.~\cite{QS_lcp} and they yield
\begin{eqnarray} 
\gamma_g^{(2)}(\omega+\lambda) =&& 
 - \frac 1{4\epsilon}(1-\epsilon) 
 (K_0g)^2  ({\mu \over - \omega})^{\epsilon}  
  \nonumber \\ 
 && - \frac 1{16\epsilon '}(N_0J)^2 
 ({\mu \over - \omega})^{2\epsilon'} 
 , 
 \label{gvertex2} 
\end{eqnarray}  
This fixes the second-order contributionn to $Z_g$,
\begin{equation} 
(Z_g-1)^{(2)} = \frac {1}{4 \epsilon} (K_0g) ^2 
 + \frac {1}{16 \epsilon'} (N_0J)^2. 
\end{equation} 
 
\begin{figure} 
\vspace{2ex} 
\centering 
\vbox{\epsfxsize=60mm\epsfbox{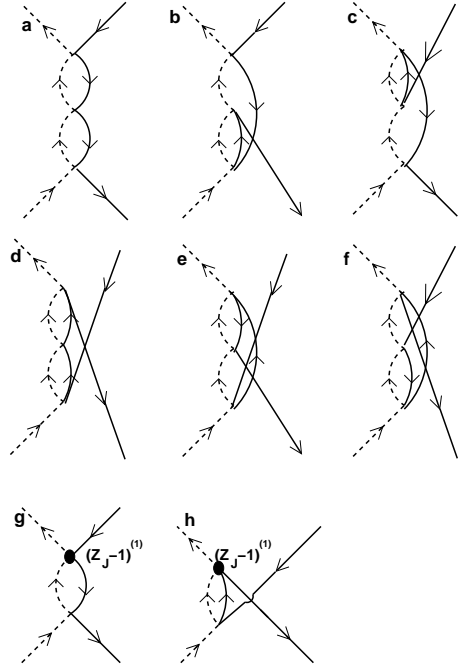}} 
\vspace{3ex} 
\caption{The vertex correction 
diagrams for the coupling $J$ to the $J^2$ order. Here,
(a)-(f) are the ``parquet'' diagrams; (g)(h) are diagrams with counterterms
$(Z_J-1)^{(1)}$.}
\label{fig:J-diagram6} 
\end{figure}

The corrections to the coupling constant $J$
include not only the direct perturbation diagrams,
but also contributions from the first order counterterm
$(Z_J-1)^{(1)}J$. Figs. 5(c)(d) in Ref.~\onlinecite{QS_lcp} give 
\begin{eqnarray} 
\gamma_J^{(2)I}(\omega+\lambda)=&&- \frac 1{4\epsilon}(1-\epsilon) 
 (K_0g)^2  ({\mu \over - \omega})^{\epsilon}  
  \nonumber \\ 
 && - \frac 1{16\epsilon '}(N_0J)^2 
 ({\mu \over - \omega})^{2\epsilon'}  . 
\end{eqnarray}
The so-called ``parquet'' diagrams as in Kondo problems, shown in
Figs.~\ref{fig:J-diagram6}(a)-(f), give 
\begin{equation} 
\gamma_J^{(2)II}(\omega+\lambda)=\frac{(N_0J)^2}{\epsilon'^2}  
({\mu \over - \omega})^{2\epsilon'} . 
\end{equation} 
The diagrams with the counterterm $(Z_J-1)^{(1)}J$, shown in
Figs.~\ref{fig:J-diagram6}(g)(h), give
\begin{eqnarray} 
\gamma_J^{(2)III}(\omega+\lambda)=&&2(Z_J-1)^{(1)} \gamma_J^{(1)}(\omega+\lambda)
  \nonumber \\ 
 =&&-\frac{2(N_0J)^2}{\epsilon'^2}  
({\mu \over - \omega})^{\epsilon'}  
 , 
\end{eqnarray} 
where the prefactor 2 is a symmetry factor.
From these contributions, we obtain the second order 
correction  to $Z_J$:
\begin{equation} 
(Z_J-1)^{(2)} = \frac {1}{4 \epsilon} (K_0g) ^2 
 + \frac {1}{16 \epsilon'} (N_0J)^2  
 + \frac {1}{ \epsilon'^2} (N_0J)^2. 
\end{equation} 

\subsection{Third order result} 
 
\begin{figure} 
\vspace{2ex} 
\centering 
\vbox{\epsfxsize=60mm\epsfbox{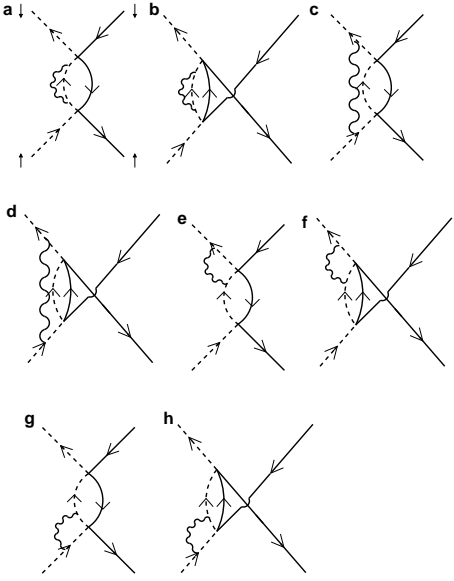}} 
\vspace{3ex} 
\caption{The direct perturbative vertex diagrams for the coupling 
$J$ to the $g^2J$ order.} 
\label{fig:J-diagram3} 
\end{figure}

\begin{figure} 
\vspace{2ex} 
\centering 
\vbox{\epsfxsize=60mm\epsfbox{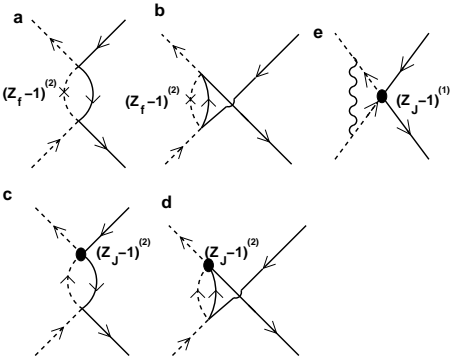}} 
\vspace{3ex} 
\caption{The vertex correction
diagrams for the coupling $J$ to the $g^2J$ order 
with counterterms $(Z_f-1)^{(2)}$, $(Z_J-1)^{(1)}$ and 
$(Z_J-1)^{(2)}$.} 
\label{fig:J-diagram4} 
\end{figure}

To the third order of perturbation, we need only to consider the $g^2J$ order
corrections to the $J$ vertices. (There are also $J^3$ order corrections
to the $J$ vertices, but these contributions are higher than $\epsilon^2$
order.) The direct perturbative 
diagrams are shown in Fig.~\ref{fig:J-diagram3}. Summing up
the contributions of these diagrams, we get 
\begin{eqnarray} 
\gamma_J^{(3)I}(\omega+\lambda) =&& 
 \left( \frac{1}{4\epsilon'}-\frac{1}{4\epsilon \epsilon '} 
-\frac{1}{\epsilon(\epsilon+\epsilon ')} \right) \nonumber \\ 
 && \times (K_0g)^2(N_0J)({\mu \over - \omega})^{\epsilon+\epsilon '}  . 
\label{gamma-J-3-1}
\end{eqnarray} 
Fig.~\ref{fig:J-diagram4} shows the diagrams with counterterms to $g^2J$ order. 
The contributions from Figs.~\ref{fig:J-diagram4}(a)(b) 
give rise to
\begin{eqnarray} 
\gamma_J^{(3)II}(\omega+\lambda)=&&-(Z_f-1)^{(2)} \gamma_J^{(1)}(\omega+\lambda)
 \nonumber \\ 
 =&& \frac{3(K_0g)^2(N_0J)}{4\epsilon \epsilon'} 
 ({\mu \over - \omega})^{\epsilon '}  
 ;  
\end{eqnarray} 
those from Figs.~\ref{fig:J-diagram4}(c)(d) yield
\begin{eqnarray} 
\gamma_J^{(3)III}(\omega+\lambda)=&&2(Z_J-1)^{(2)} \gamma_J^{(1)}(\omega+\lambda)
 \nonumber \\ 
 =&& \frac{(K_0g)^2(N_0J)}{2\epsilon \epsilon'} 
 ({\mu \over - \omega})^{\epsilon '}  
 ;  
\end{eqnarray} 
and finally, Fig.~\ref{fig:J-diagram4}(e) leads to
\begin{eqnarray} 
\gamma_J^{(3)IV}(\omega+\lambda)=&&(Z_J-1)^{(1)} \gamma_J^{(2)}(\omega+\lambda)
 \nonumber \\ 
 =&& \left( \frac 1{4\epsilon \epsilon'} -\frac{1}{4\epsilon'}\right) 
  (K_0g)^2(N_0J) 
 ({\mu \over - \omega})^{\epsilon}  
 . 
\label{gamma-J-3-4}
\end{eqnarray} 
Summing up $\gamma_J^{(3)I,II,III,IV}$ and subtracting poles, we have 
\begin{equation} 
(Z_J-1)^{(3)} = -\left( \frac {1}{\epsilon'(\epsilon+\epsilon')}
 +\frac{1}{4\epsilon \epsilon'} \right) (K_0g)^2(N_0J) .
\label{gamma-J-3}
\end{equation} 
Note that the simple poles in Eqs. (\ref{gamma-J-3-1},\ref{gamma-J-3-4})
cancel with each other, leaving only double poles in the net result,
Eq. (\ref{gamma-J-3}).
Therefore, no $g^2J$ terms will appear in the beta functions. 
 
\subsection{Fourth order result} 
We now turn to contributions of order $g^4$, the only terms we are interested
in to 
this order.
First, we consider the $f-$electron  
self-energy. Figs.~\ref{fig:f-selfenergy}(c), (d) are 
the direct perturbative contributions to $f-$electron self-energy 
of this order. Figs.~\ref{fig:f-selfenergy}(e), (f), 
on the other hand, are the diagrams with counter terms of 
order $g^2$ and $J^2$. Keeping only 
the
$g^4$ order
terms, we have: 
\begin{eqnarray}  
\Sigma_f^{(4)(\ref{fig:f-selfenergy}c+\ref{fig:f-selfenergy}d)} (\omega+\lambda) =&&  
 \omega \left( \frac {15}{32 \epsilon^2}+ \frac 3{8\epsilon} \right) (K_0g)^4  
 ({\mu \over - \omega})^{2\epsilon}, \nonumber \\  
\Sigma_f^{(4)(\ref{fig:f-selfenergy}e)} (\omega+\lambda) =&&  
   -(Z_f-1)^{(2)}\Sigma_f^{(2)}(\omega+\lambda) \nonumber \\ 
  = &&- \omega  \frac {9}{16 \epsilon^2} (K_0g)^4 ({\mu \over - \omega})^{ 
    \epsilon} \nonumber \\ 
\Sigma_f^{(4)(\ref{fig:f-selfenergy}f)} (\omega+\lambda) =&&  
   2(Z_g-1)^{(2)}\Sigma_f^{(2)}(\omega+\lambda) \nonumber \\ 
  = &&- \omega  \frac {3}{8 \epsilon^2} (K_0g)^4 ({\mu \over - \omega})^{ 
      \epsilon} 
\label{selfenergy_4}  
\end{eqnarray} 
The resulting corrections to $Z_f$ in this order is given by 
\begin{equation} 
(Z_f-1)^{(4)} = -\frac {15}{32 \epsilon^2} (K_0g) ^4 
 + \frac {3}{8 \epsilon} (K_0g)^4. 
\end{equation} 
 
\begin{figure} 
\vspace{2ex} 
\centering 
\vbox{\epsfxsize=60mm\epsfbox{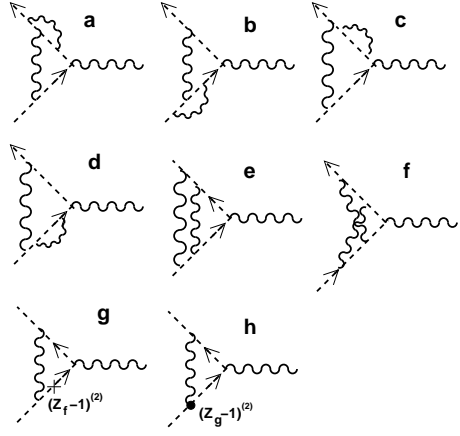}} 
\vspace{3ex} 
\caption{The vertex correction 
diagrams for the coupling $g$ to the $g^4$ order. 
(a)-(f) are the 
direct perturbative diagrams while (g), (h) are
diagrams with  counterterms} 
\label{fig:g-diagram} 
\end{figure}

Next, we consider the $g-$vertex corrections. They are specified in
Fig.~\ref{fig:g-diagram}. The contributions from  the direct 
perturbative diagrams, Figs.~\ref{fig:g-diagram}(a)-(f),
sum up to
\begin{equation} 
\gamma_g^{(4)~a-f} (\omega+\lambda)= \left( \frac 9{32\epsilon^2} - \frac 7{16\epsilon}  
 \right) (K_0g)^4 ({\mu \over - \omega})^{ 
      2\epsilon}; 
\end{equation} 
while those from the counterterms, Figs.~\ref{fig:g-diagram}(g)(h),
yield,
\begin{eqnarray} 
\gamma_g^{(4)~g,h}(\omega+\lambda)=&&3(Z_g-1)^{(2)}\gamma_g^{(2)}(\omega+\lambda)
\nonumber \\ 
&&-2(Z_f-1)^{(2)} 
\gamma_g^{(2)}(\omega+\lambda)  \nonumber \\ 
 =&& \left( -\frac 9{16\epsilon^2} + \frac 9{16\epsilon}  
 \right) (K_0g)^4 ({\mu \over - \omega})^{\epsilon}. 
\end{eqnarray} 
Therefore, the correction to $Z_g$ is given by 
\begin{equation} 
(Z_g-1)^{(4)} = \frac {9}{32 \epsilon^2} (K_0g) ^4 
 - \frac {1}{8 \epsilon} (K_0g)^4. 
\end{equation} 
 
\begin{figure} 
\vspace{2ex} 
\centering 
\vbox{\epsfxsize=60mm\epsfbox{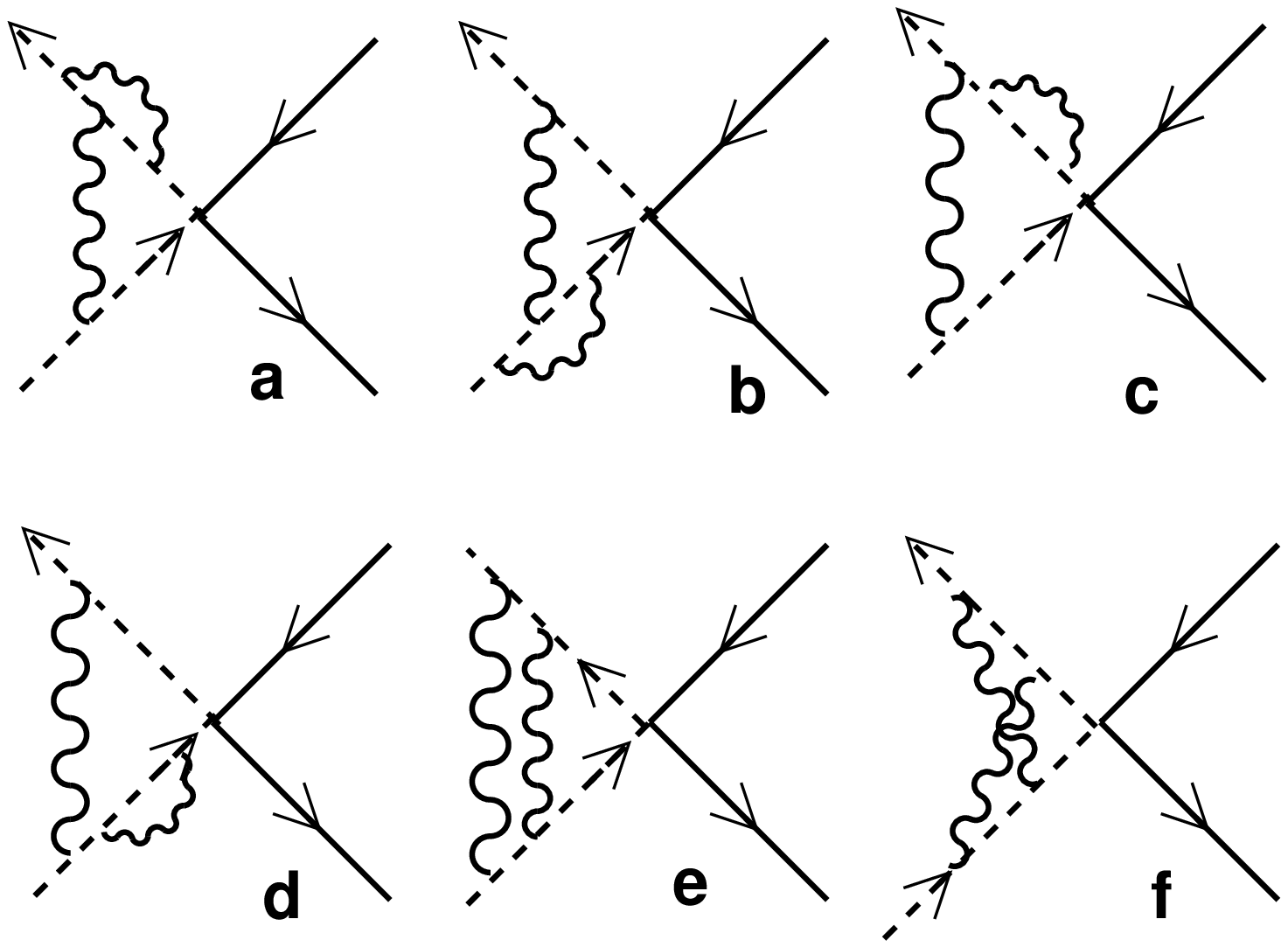}} 
\vspace{3ex} 
\caption{The vertex correction
diagrams for the coupling $J$ to the $g^4$ order.} 
\label{fig:J-diagram2} 
\end{figure} 
 
\begin{figure} 
\vspace{2ex} 
\centering 
\vbox{\epsfxsize=60mm\epsfbox{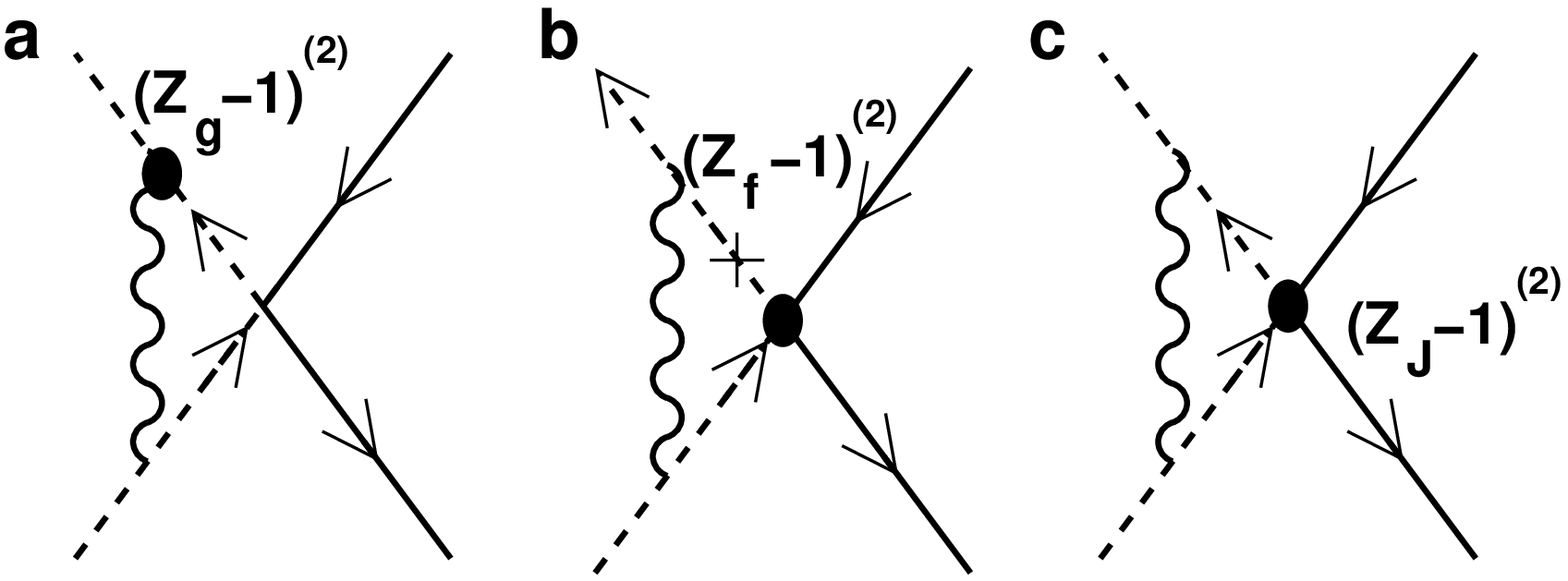}} 
\vspace{3ex} 
\caption{The vertex correction
diagrams for the coupling $J$ to the $g^4$ order with  
counterterms.} 
\label{fig:J-diagram5} 
\end{figure}

Similarly, the $J-$coupling corrections 
contain both direct perturbative diagrams (Fig.~\ref{fig:J-diagram2}) 
and diagrams with  counterterms (Fig.~\ref{fig:J-diagram5}) which
are given by 
\begin{equation} 
\gamma_J^{(4)I}(\omega+\lambda) = \left( \frac 9{32\epsilon^2} - \frac 7{16\epsilon}  
 \right) (K_0g)^4 ({\mu \over - \omega})^{ 
      2\epsilon}, 
\end{equation} 
and 
\begin{eqnarray} 
\gamma_J^{(4)II}(\omega+\lambda)=&&
\left(2(Z_g-1)^{(2)}-2(Z_f-1)^{(2)}\right)\gamma_J^{(2)} (\omega+\lambda) 
   \nonumber \\
&&+ (Z_J-1)^{(2)}\gamma_J^{(2)} (\omega+\lambda) 
   \nonumber \\ 
 =&& \left( -\frac 9{16\epsilon^2} + \frac 9{16\epsilon}  
 \right) (K_0g)^4 ({\mu \over - \omega})^{\epsilon}, 
\end{eqnarray} 
respectively. The resulting correction to $Z_J$ is  
\begin{equation} 
(Z_J-1)^{(4)} = \frac {9}{32 \epsilon^2} (K_0g) ^4 
 - \frac {1}{8 \epsilon} (K_0g)^4. 
\end{equation} 
 
We can now collect the renormalization factors $Z_f$, $Z_g$, and
$Z_J$ to the orders of our interest. The results are given in
Eq.~(\ref{RF2}).
 
\subsection{The beta functions} 
After obtaining the renormalization factors, we can calculate the beta
functions, defined in Eq.~(\ref{beta-definition}).

Taking the $\mu$ derivative of 
Eqs.~(\ref{RF_def1},\ref{RF_def2}) 
at the fixed bare couplings, we have
\begin{eqnarray}
0 = && \frac{\beta(g)}{g} + \beta(g)
       \frac{\partial \ln (Z_f^{-1}Z_g)}{\partial g}
 	\nonumber \\
	&&+ \beta(J) \frac{ \partial \ln (Z_f^{-1}Z_g)}{\partial J}
		+ \frac{\epsilon}{2} ,\nonumber \\
0=&&\frac{\beta(J)}{J} + \beta(g) 
	\frac{\partial \ln (Z_f^{-1}Z_J)}{\partial g}
	\nonumber \\
	&&+ \beta(J) \frac{ \partial \ln (Z_f^{-1}Z_{J})}{\partial J}
		+ \epsilon'.
\end{eqnarray}

After some algebra, and setting $\epsilon'=0$, we obtain the beta
functions of $g$ and $J$ as in Eq.~(\ref{rg-equations}).

\subsection{Renormalizability}

We close this appendix by commenting on
two issues regarding renormalizability. 
Our perturbative results for the self-energy and vertex corrections
as well as the renormalization factors 
contain double poles such as $1/\epsilon^2$,
$1/(\epsilon')^2$ and 
$1/\epsilon \epsilon'$.
However, we find that terms of the form
$\frac 1{\epsilon} \ln \frac{\mu}{-\omega}$
all cancel out
after we carry out expansions such as 
\begin{eqnarray}
({\mu \over - \omega})^{\epsilon} = 1 + \epsilon  
  \ln {\mu \over - \omega} + O(\epsilon^2) . 
\label{expansion}
\end{eqnarray}
This is in accordance with the requirement of renormalizability
\cite{Zinn-Justin,Brezin,Gross,tHooft}.
In addition, we find that the double poles in the renormalization
factors cancel out in the beta functions. 
This is also a consistency check for the minimal subtraction procedure,
and is in accordance with the requirement that 
beta functions are analytic.

\section{Calculations of the local spin susceptibility}
\label{sec:appen-chiloc}

\subsection{The local susceptibility to order $\epsilon^2$}

The local spin correlation function is defined in Eq.~(\ref{chi_loc_define}). 
We find it convenient to carry out our calculation in $\tau$ space.
(The self-energy and vertex correction diagrams in
Appendix \ref{sec:appen-rg} were calculated
in $\omega$ space.)
So we take the 
expressions for the boson and 
conduction electron
propagators as 
\begin{eqnarray}
G_{\phi}^0(\tau)=&& {\tilde K_0}^2
\left ( \frac {\pi/\beta}{\sin{{\pi\tau} \over \beta}}
\right )^{2-\epsilon}
 \nonumber \\
G_{c}^0(\tau) = && {\tilde N_0}
\left ( \frac {\pi/\beta}{\sin{{\pi\tau} \over \beta}}
\right )^{1-\epsilon '}
\end{eqnarray}
where  ${\tilde K_0}$ and ${\tilde N_0}$ can be determined 
from $K_0$ and $N_0$ by
\begin{eqnarray}
{\tilde K_0}^2=K_0^2 \Gamma(2-\epsilon) \nonumber \\
{\tilde N_0}^2=N_0^2 \frac{\Gamma(2-2\epsilon')}{\Gamma^2(1-\epsilon')},
\end{eqnarray}
where $\Gamma(x)$ is the Euler Gamma function.

We now calculate the bare spin susceptibility
in terms of the bare Hamiltonian
(instead of the renormalized Hamiltonian used in Appendix~\ref{sec:appen-rg}).
The zero-th order result is given by
\begin{eqnarray}
\tilde {\chi}_B^{(0)}(\tau)&& =
 -\frac 12 \int_0^{\beta} d\tau~G_f^0(\tau)G_f^0(-\tau) 
  = \frac 12~ e^{-\beta \lambda} \nonumber \\   
\chi^{(0)}_B(\tau) && = \lim_{\lambda \rightarrow \infty} 
 {1 \over 2}~ e^{\beta \lambda}~
 \tilde {\chi}_B(\tau)   = \frac 14,
\end{eqnarray}
where $G_f^0(\tau)$ is the bare $f-$electron propagator. This result is 
simply $S(S+1)/3$ with $S=1/2$. 

To the order $g^2$ and $J^2$,
the diagrams have been shown in Fig. 8 of 
Ref.~\cite{QS_lcp}. Using the dimensional regularization
adopted here, we find
\begin{eqnarray}
\chi^{(2)}_B(\tau) =&& -{1 \over 4}
     \frac {2(K_0g_B)^2  \tau^{\epsilon}}{\epsilon} 
     \left(1+ C \epsilon + \frac 12 
	 (C^2+ \frac{\pi^2}{6}) \epsilon^2 
	 \right) \nonumber \\
&& -{1 \over 4} \frac {(N_0J_B)^2
 \tau^{2\epsilon '}}{2\epsilon '}
 (1+\frac {\pi^2}{6} \epsilon'^2) ,
\label{chi_loc_result2}
\end{eqnarray}
where $C=0.577215 \cdot \cdot \cdot $ is the Euler Gamma constant.

\begin{figure}
\vspace{2ex}
\centering
\vbox{\epsfxsize=60mm\epsfbox{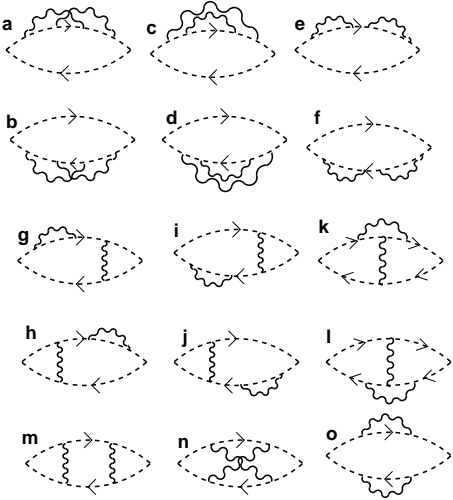}}
\vspace{3ex}
\caption{Diagrams for the local spin susceptibility
to the $g^4$ order.}
\label{fig:chi-diagram}
\end{figure}

To $g^4$ order, the diagrams are shown in
Fig.~\ref{fig:chi-diagram}. (Structurally, they are similar
to those considered in ref.~\onlinecite{Vojta}.) The result is
\begin{eqnarray}
\chi^{(4)}_B(\tau) =&& {1 \over 4}
 \left( \frac{4}{\epsilon^2}+
\frac{1+8C}{\epsilon} + 2 +2C+ 8C^2
 + \frac{\pi^2}{3} \right) \times \nonumber \\
 &&\times (K_0g_B)^4 \tau^{2\epsilon}
\end{eqnarray} 

We now express $\chi_B$ in terms of the renormalized coupling constants,
$g$ and $J$, by using Eqs.~(\ref{RF_def1},\ref{RF_def2}) in combination
with our results for the renormalization factors, Eq. (\ref{RF2}).
We can then determine the renormalization factor $Z$, defined in
Eq. (\ref{Z-defined}), by demanding that all the pole terms in
$\chi_B$ are removed by those of $Z^{-1}$ so that the renormalized
susceptibility, $\chi=Z^{-1}\chi_B$, is regular. The result is
\begin{eqnarray}
Z^{-1}=&&1+\frac{2 (K_0g)^2}{\epsilon} + \frac{ (N_0J)^2}{2\epsilon '}.
\nonumber \\
 &&+\frac {4(K_0g)^4}{\epsilon^2}- \frac {(K_0g)^4}{\epsilon}.
\label{RFZ_inverse}
\end{eqnarray}
Eq.~(\ref{RF3}) then follows.

The renormalized susceptibility, $\chi=Z^{-1}\chi_B$,
is equal to
\begin{equation}
\chi(\tau)=\chi^{(0)} \left[ A(g, J, \epsilon)- 
 B(g, J, \epsilon) \ln (\mu \tau) 
 + O(\ln ^2(\mu \tau)) \right],
\end{equation}
where, to the $\epsilon ^2$ order,
\begin{eqnarray}
A(g, J, \epsilon) =&&
 1-2(K_0g)^2(C+\frac 12 
	 (C^2+ \frac{\pi^2}{6}) \epsilon) \nonumber \\
&&+(K_0g)^4(2 + 2C + 4C^2 - \frac{\pi^2}{3}) \\	 
B(g, J, \epsilon)=&& 2(K_0g)^2 (1+C \epsilon) +(N_0J)^2 
 -2(K_0g)^4(1+4C). \nonumber 
\end{eqnarray}
At the critical point(``C'' in Fig.~\ref{fig:RGflow_e01}),
 the spin correlation has the form
\begin{equation}
  \chi(\tau)=\chi^{(0)}  A(g^*, J^*, \epsilon)\frac 1{(\mu \tau)^{\eta}},
\end{equation}
where the exponent $\eta$ can either be calculated from the renormalization 
factor $Z$ in Eq.~(\ref{eta-expression}), or by 
$B(g^*, J^*, \epsilon)/A(g^*, J^*, \epsilon)$,
which also yields $\eta=\epsilon$.
The renormalized amplitude factor
$A(g^*, J^*, \epsilon)$ is given by
\begin{equation}
  A(g^*, J^*, \epsilon)=
   1- C \epsilon 
      + \left( \frac12 + \frac14 C + \frac12 C^2 - 
            \frac16 \pi^2 \right)  \epsilon^2.
\end{equation}
For large values of $\epsilon$, it would require an appropriate 
resummation such as Pad\'e approximation.

For the local moment fixed point, the local spin correlation has the form
\begin{eqnarray}
\chi(\tau)=&&\chi^{(0)} \left[ A_L(\epsilon)- 
 B_L(\epsilon) \ln (\mu \tau) + O(\ln ^2(\mu \tau)) \right], \nonumber \\
A_L(\epsilon) =&&A(g, J, \epsilon) \vert_{g=g^*_L, J=J^*_L=0} \nonumber \\
=&& 1- C \epsilon 
      + \left( \frac12  + \frac12 C^2 - 
            \frac16 \pi^2 \right)  \epsilon^2, \nonumber \\
B_L(\epsilon)=&& B(g, J, \epsilon) \vert_{g=g^*_L, J=J^*_L=0} \nonumber \\
=&& \epsilon -C \epsilon^2.
\end{eqnarray}
Again, $B_L(\epsilon)/A_L(\epsilon)=\epsilon$ to the $\epsilon^2$ order.

\subsection{The local susceptibility to all orders in $\epsilon$}

\begin{figure}
\vspace{2ex}
\centering
\vbox{\epsfxsize=60mm\epsfbox{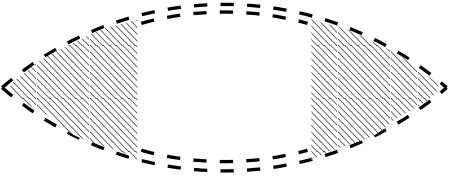}}
\vspace{3ex}
\caption{The full local spin correlation function.
Each double line represents 
the full $f-$electron propagator, while a shaded area denotes the full vertex.}
\label{fig:Full_spin_cor}
\end{figure}

We now discuss the contributions to $\chi$, to all orders 
of the perturbation theory. We can group all the diagrams
in a manner illustrated in Fig.~\ref{fig:Full_spin_cor}. 
Here, each double line represents the full $f-$electron propagator, while a 
shaded area denotes the full vertex. The renormalization factor for the spin,
$Z$, is the product of the renormalization factors for the
full f-electron propagators and those for the vertices.
The renormalization factor for a full f-electron propagator is
equal to $Z_f$. By inspecting the diagrams for each vertex,
it is straightforward to see that to each order, 
this vertex is identical to that of the $g-$vertex.
As a result, each vertex contributes a factor $Z_g^{-1}$.
Taking these two things together, we end up with
Eq. (\ref{RF_relation}), to all orders.
As discussed in Section \ref{sec:suscep2},
Eq.~(\ref{RF_relation}) then leads to
Eqs.~(\ref{eta-epsilon},\ref{critical-exp}).

\section{RG analysis for anisotropic cases}
\label{sec:appen-aniso}

In the anisotropic case, the diagrams are topologically the same as
those in the isotropic case, except for different coupling vertices.
In the following, we will simply
list the results.

The renormalization factors are given as    
\begin{eqnarray}
Z_f &=& 1 - {1 \over \epsilon} \left( \frac{(K_0 g_{\perp})^2}{2}
	+ \frac{(K_0 g_{z})^2}{4} \right)  \nonumber \\
    &-& {1 \over \epsilon^2 } \left( \frac{(K_0 g_{\perp})^4 } {8}
	+ \frac{3(K_0 g_{\perp})^2 (K_0 g_z)^2}8
	- \frac{(K_0 g_z)^4}{32} \right) \nonumber \\
    &+& {1 \over \epsilon} \left( \frac{(K_0 g_{\perp})^4}8
	+ \frac{(K_0 g_{\perp})^2 (K_0 g_z)^2}4 \right) \nonumber \\
    &-& {1 \over \epsilon'} \left( \frac{(N_0J_{\perp})^2}8
	+ \frac{(N_0J_z)^2}{16} \right), 
\label{Zf_aniso}
\end{eqnarray}
\begin{eqnarray}
Z_{g_{\perp}} &=& 1 + { (K_0g_z)^2 \over 4\epsilon}  \nonumber \\
    &+& {1 \over \epsilon^2 } \left( 
	 \frac{(K_0 g_{\perp})^2 (K_0 g_z)^2}4
	+ \frac{(K_0 g_z)^4}{32} \right) \nonumber \\
     &-& \frac{(K_0 g_{\perp})^4}{8\epsilon} 
    + {(N_0J_z)^2 \over 16\epsilon'},
\label{Zgperp_aniso}
\end{eqnarray}
\begin{eqnarray}
Z_{g_z} &=& 1 + {1 \over \epsilon} \left( \frac{(K_0 g_{\perp})^2}{2}
	- \frac{(K_0 g_{z})^2}{4} \right)  \nonumber \\
    &&+ {1 \over \epsilon^2 } \left( \frac{3(K_0 g_{\perp})^4}8
	- \frac{(K_0 g_{\perp})^2 (K_0 g_z)^2}8
	+ \frac{(K_0 g_z)^4}{32} \right) \nonumber \\
    &&+ {1 \over \epsilon} \left( \frac{(K_0 g_{\perp})^4}8
	- \frac{(K_0 g_{\perp})^2 (K_0 g_z)^2}4 \right) \nonumber \\
    &&+ {1 \over \epsilon'} \left( \frac{(N_0J_{\perp})^2}8
	- \frac{(N_0J_z)^2}{16} \right), 
\label{Zgz_aniso}
\end{eqnarray}
\begin{eqnarray}
Z_{J_{\perp}} &=& 1 - {N_0J_z \over \epsilon '} \nonumber \\
&&+ {(K_0g_z)^2 \over 4\epsilon} + {(N_0J_z)^2 \over 16\epsilon '}
 + \frac{(N_0J_{\perp})^2+(N_0J_{z})^2}{2\epsilon'^2}
 \nonumber \\
&&-\frac{(K_0g_{\perp})^2(N_0J_{z})}{\epsilon'(\epsilon+\epsilon')}
-\frac{(K_0g_{z})^2(N_0J_{z})}{4\epsilon \epsilon'} \nonumber\\
    &&+ {1 \over \epsilon^2 } \left( 
	 \frac{(K_0 g_{\perp})^2 (K_0 g_z)^2}4
	+ \frac{(K_0 g_z)^4}{32} \right)  \nonumber \\
    && 	- \frac{(K_0 g_{\perp})^4}{8\epsilon}
\label{ZJperp_aniso}
\end{eqnarray}
\begin{eqnarray}
Z_{J_z} =&& 1 - {N_0(J_{\perp}^2/J_z) \over \epsilon '} 
 + {1 \over \epsilon} \left( \frac{(K_0 g_{\perp})^2}{2}
	- \frac{(K_0 g_{z})^2}{4} \right)  \nonumber \\
&&+ {1 \over \epsilon'} \left( \frac{(N_0J_{\perp})^2}8
	- \frac{(N_0J_z)^2}{16} \right) 
+ \frac{(N_0J_{\perp})^2}{\epsilon'^2} \nonumber \\ 
 &&- \frac{(K_0g_z)^2(N_0J_z)}{\epsilon'(\epsilon+\epsilon')}
 \nonumber \\ 
 &&- \frac{ \left( 2(K_0g_{\perp})^2-(K_0g_z)^2 \right) 
(N_0J_{\perp}^2/J_z)}{4 \epsilon \epsilon'} \nonumber\\
&&+ {1 \over \epsilon^2 } \left( \frac{3(K_0 g_{\perp})^4}8
	- \frac{(K_0 g_{\perp})^2 (K_0 g_z)^2}8
	+ \frac{(K_0 g_z)^4}{32} \right) \nonumber \\
&&+ {1 \over \epsilon} \left( \frac{(K_0 g_{\perp})^4}8
	- \frac{(K_0 g_{\perp})^2 (K_0 g_z)^2}4 \right) 
\label{ZJz_aniso}
\end{eqnarray}
Eqs.~(\ref{Zf_aniso})-
(\ref{ZJz_aniso}) reduce to the isotropic results,
Eq.~(\ref{RF2}), when we set 
$g_z=g_{\perp}$, $J_z=J_{\perp}$.
Similar to the isotropic
case, we obtain the beta functions for each coupling constant from these
renormalization factors, which are
given
in Eqs.~(\ref{beta_gperp_aniso})-
(\ref{beta_Jz_aniso}).
We can then set $g_z=0$ ($g_{\perp}=0$) to
discuss the xy (Ising) case.
 
We now turn to the calculation of the local spin susceptibility,
focusing on the xy case only.
Like in the isotropic case, we use the bare Hamiltonian.
The contributions from the perturbative diagrams
to the $g^2$, $J^2$, and $g^4$
orders are
\begin{eqnarray}
 \chi_B ^{+-(2)} (\tau) =&& 
  -\frac 14\frac{(K_0 g_{\perp}^B)^2 +(K_0g_z^B)^2}{\epsilon} \tau ^{\epsilon} 
 \nonumber\\
&&\times     \left(1+ C \epsilon + \frac 12 
	 (C^2+ \frac{\pi^2}{6}) \epsilon^2 
	 \right) \nonumber \\
  &&- \frac{(N_0 J_{\perp}^B)^2 + (N_0 J_z^B)^2}{16\epsilon '} 
  \tau^{2\epsilon '}
 (1+\frac {\pi^2}{6} \epsilon'^2) ,
\\
  \chi_B ^{+-(4)} (\tau) =&&  \frac{1}{4}\frac{(K_0 g_{\perp}^B)^4
  + (K_0 g_{\perp}^B)^2 (K_0 g_z^B)^2}2 \tau^{2 \epsilon} \times
\nonumber \\
  &&\times 
 \left( \frac{4}{\epsilon^2}+
\frac{1+8C}{\epsilon} + 2 +2C+ 8C^2
 + \frac{\pi^2}{3} \right) . \nonumber
\end{eqnarray}
(There are no contributions at the $J$ or $g^2J$ order.)
Following the same RG procedure as in Appendix~\ref{sec:appen-chiloc},
we can calculate the renormalization factor $Z$: 
\begin{eqnarray}
Z&=& 1 - {1 \over \epsilon} \left( (K_0 g_{\perp})^2
	+(K_0 g_{z})^2 \right)  \nonumber \\
 &+& {1 \over \epsilon} \left( \frac{(K_0 g_{\perp})^4}2
	+ \frac{(K_0 g_{\perp})^2 (K_0 g_z)^2}2 \right) \\
\label{Z_aniso}
 &-& {1 \over \epsilon'} \left( \frac{(N_0J_{\perp})^2}4
	+ \frac{(N_0J_z)^2}{4} \right), \nonumber 
\end{eqnarray} 
and the anomalous dimension $\eta$ at the critical point:
\begin{eqnarray}
\eta &=&  \frac{ d \log Z}{d \log \mu} \vert _{g_i=g_i^*, J_i=J_i^*}
          \nonumber \\
     &=& (K_0g_{\perp}^*)^{2} +(K_0g_{z}^*)^{2}
        -   (K_0 g_{\perp}^*)^{4} \nonumber \\
     &-& (K_0 g_{\perp}^*)^{2} (K_0 g_z^*)^{2}
      +  \frac{(N_0J_{\perp}^*)^{2}}2 + \frac{(N_0J_z^*)^{2}}{2}
\label{crit_exp_ansio}
\end{eqnarray}

Setting $g_z=0$ (for the xy case), the exponent becomes
\begin{equation}
 \eta=(K_0 g_{\perp}^*)^{2} -   (K_0 g_{\perp}^*)^{4} 
      +  \frac{(N_0J_{\perp}^*)^{2}+(N_0J_z^*)^{2}}2 .
\label{crit_exp_ansio_xy}
\end{equation}
The renormalized amplitude $A(g,J,\epsilon)$ is given by
\begin{eqnarray}
A(g, J, \epsilon) =&&
 1-(K_0g_{\perp})^2(C+\frac 12 
	 (C^2+ \frac{\pi^2}{6}) \epsilon) \nonumber \\
&&+(K_0g_{\perp})^4(1 + C + 2C^2 - \frac{\pi^2}{6}) 	 
\end{eqnarray}

At the critical point, taking account of Eq.~(\ref{fp-xy}), we find 
\begin{eqnarray}
\eta&&= \epsilon, \nonumber \\
A_C(\epsilon)&&=1- C \epsilon 
      + \left( 1  +\frac 38 C + \frac32 C^2 - 
            \frac14 \pi^2 \right)  \epsilon^2.
\end{eqnarray}

Near the bosonic stable fixed point where $J_{\perp}^*=J_z^*=0$,
which is given by Eq.~(\ref{fp-xy-lm}), the results are
\begin{eqnarray}
 \eta_L=&&(K_0 g_{\perp}^*)^{2} -   (K_0 g_{\perp}^*)^{4} \nonumber \\
 =&& \epsilon , \nonumber \\
A_L(\epsilon)=&&1- C \epsilon 
      + \left( 1  + \frac32 C^2 - 
            \frac14 \pi^2 \right)  \epsilon^2.
\label{crit_exp_ansio_xy_lm}
\end{eqnarray}

\section{Kink-gas analysis of the Ising case}
\label{sec:appen-ising}

We now turn to the Ising case, by setting $g_{\perp}=0$.
The mapping to the kink-gas action is similar to that given in
ref.~\onlinecite{Smith1}. The resulting action, describing
a one-dimensional
long-ranged
statistical-mechanical model,
is similar to Eq. (7) of ref.~\onlinecite{Smith1} and,
in our notation, takes the following form:
\begin{eqnarray}
S(\tau_{2n}, &&...,\tau_{1})= - 2n ~\ln(y_j) 
+\sum_l (-1)^l ~h ~({\tau_{l+1} - \tau_l})/
\tau_0
\nonumber\\
&& + \sum_{l < m}(-1)^{l+m} [ 2\kappa_j
\ln ({\tau_{m}-\tau_{l}})/
{\tau_{0}} 
+ K({\tau_{m}-\tau_{l}}) ]
\label{kinkgas}
\end{eqnarray}
where $[\tau_{2n},...,\tau_{1}]$, for $n=1, 2, ...$, labels a sequence
of spin flips (kinks) along the imaginary time axis,
and $y_j$ and $\kappa_j$ are defined in Eq.~(\ref{RG-charges-ising}).
The last term, $K(\tau)$, originates from the $g_z-$coupling and takes
the form:
\begin{eqnarray}
K(\tau) = \kappa_g [(\tau / \tau_0)^\epsilon - 1 ] /\epsilon
\label{K}
\end{eqnarray}
where $\kappa_g$ is also specified by Eq.~(\ref{RG-charges-ising}).
The RG equations for this kink-gas problem has already been derived in
ref.~\onlinecite{Smith1} and are given as follows:
\begin{eqnarray}
\beta (y_j) ~&&=~ - y_j ( 1 - \kappa_j - \kappa_g/2 )\nonumber\\
\beta (\kappa_j) ~&&=~ 4 \kappa_j y^2 \nonumber\\
\beta (\kappa_g) ~&&=~ - \kappa_g ( \epsilon - 4 y_j^2 ) \nonumber\\
\beta (h) ~&&=~ -h ( 1 - 2 y_j^2)
\label{scaling-ising}
\end{eqnarray}
This procedure is valid for arbitrary values of the stiffness 
constants $\kappa_j$ and $\kappa_g$,
provided the fugacity $y_j$ is small.

An unstable fixed point still exists. To the linear order in $\epsilon$,
it is located at 
\begin{eqnarray}
(y_j^* )^2 && = \epsilon / 4 \nonumber\\
\kappa_g^* && = 2 \nonumber\\
\kappa_j^* && = 0
\label{unstable-fp-ising}
\end{eqnarray}

From Eq. (\ref{scaling-ising}), we can easily see that $\kappa_g$ is
in fact irrelevant around this fixed point. We can then
determine the critical exponents by staying within the 
$y_j-\kappa_g$ plane: the projection of the RG flow on the 
$y_j-\kappa_g$ plane is given in Fig.~\ref{fig:Ising_RGflow}.
Within the $\epsilon-$expansion, then, the critical exponents of 
the unstable fixed point become identical to their counterparts
for a similar fixed point of the classical ferromagnetic
Ising chain with an long-range interaction that decays in distance
in terms of a power-law exponent $2-\epsilon$\cite{Kosterlitz}.
The critical exponent $\eta$, for instance, can be straightforwardly
calculated by combing Eq.~(\ref{scaling-ising}) and 
Eq.~(\ref{unstable-fp-ising}). The result is $\eta=\epsilon$,
the same value as it takes at the critical point of the
corresponding long-ranged Ising chain\cite{Kosterlitz}.


\begin{references}

\bibitem{StewartRMP} G.\ R.\ Stewart,
Rev.\ Mod.\ Phys.\ {\bf 73}, 797 (2001).

\bibitem{vonLohneysen}
H.\ v.\ L\"ohneysen {\it et al.},
Phys.\ Rev.\ Lett.\ {\bf 72}, 3262 (1994);
H.\ v.\ L\"ohneysen, J.\ Magn.\ Magn.\ Mater.\ {\bf 200}, 532 (1999).

\bibitem{Lonzarich}  N.\ D.\ Mathur {\it et al.},
Nature {\bf 394}, 39 (1998);
F. M. Grosche {\it et al.},
J. Phys.: Condens. Matter {\bf 13}, 2845 (2001).

\bibitem{Steglich} O.\ Trovarelli {\it et al.},
Phys.\ Rev.\ Lett.\ {\bf 85}, 626 (2000).

\bibitem{Stewart} K.\ Heuser {\it et al.},
Phys.\ Rev.\ B {\bf 57}, R4198 (1998).

\bibitem{Flouquet} S.\ Raymond, L.\ P.\ Regnault, J.\ Flouquet,
A.\ Wildes, and P.\ Lejay, J. Phys.: Condens. Matter {\bf 13},
8303 (2001).

\bibitem{Thompson} V. A. Sidorov {\it et al.},
cond-mat/0202251.

\bibitem{deVisser} P.\ Estrela, A.\ de Visser, T.\ Naka,
F.\ R.\ de Boer, and
L.\ C.\ J.\ Pereira, cond-mat/0009324.

\bibitem{Maple} C. L. Seaman {\it et al.},
Phys.\ Rev.\ Lett.\ {\bf 67}, 2883 (1991);
M.\ B.\ Maple {\it et al.},
J.\ Low Temp.\ Phys.\ {\bf 95}, 225 (1994).

\bibitem{Andraka} B.\ Andraka and A.\ M.\ Tsvelik,
Phys.\ Rev.\ Lett.\ {\bf 67}, 2886 (1991).

\bibitem{Aronson} M.\ C.\ Aronson {\it et al.},
Phys.\ Rev.\ Lett.\ {\bf 75}, 725 (1995).

\bibitem{MacLaughlin} O.\ O.\ Bernal, D.\ E.\ MacLaughlin,
H.\ G.\ Lukefahr, and B.\ Andraka,
Phys.\ Rev.\ Lett.\ {\bf 75}, 2023 (1995).

\bibitem{Schroder2} A.\ Schr\"{o}der {\it et al.},
Nature {\bf 407}, 351 (2000).

\bibitem{Schroder1} A.\ Schr\"oder, G.\ Aeppli, E.\ Bucher,
R.\ Ramazashvili, and P.\ Coleman,
Phys.\ Rev.\ Lett.\ {\bf 80}, 5623 (1998).

\bibitem{Stockert} O.\ Stockert, H.\ v.\ L\"{o}hneysen, A.\ Rosch,
N.\ Pyka, and M.\ Loewenhaupt,
Phys.\ Rev.\ Lett.\ {\bf 80}, 5627 (1998).

\bibitem{Stockert2}
A.\ Rosch, A.\ Schr\"oder, O.\ Stockert, and H.\ v.\ L\"ohneysen,
Phys.\ Rev.\ Lett. {\bf 79}, 159 (1997).

\bibitem{QS_nature} Q.\ Si, S.\ Rabello, K.\ Ingersent, and J.\ L.\ Smith,
Nature {\bf 413}, 804 (2001).

\bibitem{QS_lcp} Q.\ Si, S.\ Rabello, K.\ Ingersent, and J.\ L.\ Smith,
cond-mat/0202414.

\bibitem{IJMPB} Q.\ Si, J.\ L.\ Smith, and K.\ Ingersent,
Int.\ J.\ Mod.\ Phys.\ B {\bf 13}, 2331 (1999).

\bibitem{Review} For a recent review on related issues 
see P.\ Coleman, C.\ P\'{e}pin, Q.\ Si, and R.\ Ramazashvili,
J.\ Phys.: Condens.\ Matter {\bf 13}, R723 (2001).

\bibitem{Smith1} Q.\ Si and J.\ L.\ Smith,
Phys.\ Rev.\ Lett.\ {\bf 77}, 3391 (1996).

\bibitem{Smith2} J.\ L.\ Smith and Q.\ Si,
cond-mat/9705140;
Europhys.\ Lett.\ {\bf 45}, 228 (1999).

\bibitem{Sengupta} A.\ M.\ Sengupta,
cond-mat/9707316;
Phys.\ Rev.\ B {\bf 61}, 4041 (2000).

\bibitem{Vojta} M.\ Vojta, C.\ Buragohain, and S.\ Sachdev,
Phys.\ Rev.\ B {\bf 61}, 15152 (2000).

\bibitem{Sachdev-Ye} S. Sachdev and J. Ye, Phys. Rev. Lett.
{\bf 70}, 3339 (1993).

\bibitem{GPS} A. Georges, O. Parcollet, S. Sachdev,
Physical Review B 63, 134406 (2001).

\bibitem{Hewson} A.\ C.\ Hewson, {\it The Kondo Problem to Heavy Fermions}
(Cambridge Univ.\ Press, Cambridge, 1993).

\bibitem{Abrikosov} A.\ A.\ Abrikosov,
Physics {\bf 2}, 5 (1965).

\bibitem{Zinn-Justin} J. Zinn-Justin,
{\it Quantum Field Theory and Critical Phenomena}, 
(Oxford University Press, London, 1996),
Chap. 11.

\bibitem{Brezin}
E.\ Br\'ezin, J.\ C.\ Le\ Guillou, and J.\ Zinn-Justin, in 
{\it Phase Transitions and Critical Phenomena}, edited by
C.\ Domb and M.\ S.\ Green (Academic, London, 1976), Vol.\ 6. 

\bibitem{Gross} D. J. Gross, in
{\it Methods in Field Theory},
edited by R. Balian and J. Zinn-Justin,
(North-Holland Publishing, Amsterdam, 1975),
p. 141.

\bibitem{tHooft} G. 't Hooft, Nucl. Phys. B{\bf 61},
455 (1973); G. 't Hooft and M. Veltman,
{\it ibid.} {\bf 44}, 189 (1972).

\bibitem{footnote} We adopt the sign convention of 
the field-theory approach.

\bibitem{footnote2} Here we speak of phases  having in mind 
the one-dimensional long-ranged statistical-mechanical model
associated with the Bose-Fermi Kondo model.

\bibitem{Fisher} M.\ E.\ Fisher, S.-K.\ Ma, and B.\ G.\ Nickel,
Phys.\ Rev.\ Lett.\ {\bf 29}, 917 (1972).

\bibitem{Suzuki} M.\ Suzuki, Prog.\ Theor.\ Phys.\ {\bf 49},
424, 1106, 1440 (1973).

\bibitem{Spohn} H. Spohn and R. Dumche, J. Stat. Phys. {\bf 41}, 389 (1985)
and references therein.

\bibitem{Smith3} J.\ L.\ Smith and Q.\ Si,
Phys.\ Rev.\ B {\bf 61}, 5184 (2000).

\bibitem{Chitra} R.\ Chitra and G.\ Kotliar,
Phys.\ Rev.\ Lett.\ {\bf 84}, 3678 (2000).

\bibitem{Georges} A.\ Georges, G.\ Kotliar, W.\ Krauth, and
M.\ J.\ Rozenberg, Rev.\ Mod.\ Phys.\ {\bf 68}, 13 (1996).

\bibitem{Kosterlitz} J.\ M.\ Kosterlitz,
Phys.\ Rev.\ Lett.\ {\bf 76}, 1577 (1976).

\bibitem{aps} L.\ Zhu and Q.\ Si, Bull. Am. Phys.
Soc. {\bf 47}(1), 89 (2002)[March].

\bibitem{Zarand} G. Zarand, private communication.

\end{references}
\end{document}